# Which Factors Matter Most?
# Can Startup Valuation be Micro-Targeted?[1]

Max Berre[2]

**ABSTRACT**

While startup valuations are influenced by revenues, risks, age, and macroeconomic conditions, specific causality is traditionally a black box. Because valuations are not disclosed, roles played by other factors (industry, geography, and intellectual property) can often only be guessed at. VC valuation research indicates the importance of establishing a factor-hierarchy to better understand startup valuations and their dynamics, suggesting the wisdom of hiring data-scientists for this purpose. Bespoke understanding can be established via construction of hierarchical prediction models based on decision trees and random forests. These have the advantage of understanding which factors matter most. In combination with OLS, the also tell us the circumstances of when specific causalities apply. This study explores the deterministic role of categorical variables on the valuation of start-ups (i.e. the joint-combination geographic, urban, and sectoral denomination-variables), in order to be able to build a generalized valuation scorecard approach. Using a dataset of 1,091 venture-capital investments, containing 1,044 unique EU and EEA, this study examines microeconomic, sectoral, and local-level impacts on startup valuation. In principle, the study relies on Fixed-effects and Joint-fixed-effects regressions as well as the analysis and exploration of divergent micro-populations and fault-lines by means of non-parametric approaches combining econometric and machine-learning techniques.

*Keywords* Valuation, Startup Valuation, Venture Capital, Entrepreneurial Finance, Machine Learning, Hierarchical Analysis.

[1] This article has been sponsored by the Partners of the "Finance for innovation" Chair at Audencia Business School, especially Early Metrics and Sowefund, who provided financial support to Max Berre during his PhD course. The author wishes to express sincere gratitude for this support.
[2] PhD Candidate, Audencia Business School, Nantes, France and Université de Lyon, iaelyon, Magellan, Lyon, France
mberre@audencia.com

# 1. Introduction

Why do startups in California attract higher valuations than those in New York? Or ones based in London attract higher valuations than those in Paris, Berlin, or Milan, even when based in similarly-sized economies, sharing the same industries and many of the same investors? What drives this? Which factors matter most? While classical economic theory describes valuations as being based on revenues, growth-rates, and risk-adjusted discount-rates, valuation of startups proves the exception to the rule.

Given their opacity, short histories, and vast array of intangible assets, startups are notoriously difficult to value (Damodaran, 2009). This has given rise to a diversity of valuation approaches dependent on drivers known to have valuation-impacts on early-stage startups in various phases. Valuation-approaches such as discounted-cashflow (DCF), multiples-valuation, and scorecard-valuation rely on inputs such as assets or performance-measures, while other strands of the literature describe the impact of market-characteristics and competitive-environment on firm-value. Adding to the cacophony, widespread press-coverage, describes dramatic valuation-divergences along geographic and industry lines– divergences not wholly explained by growth, risk, revenue, or assets–.

Scarcity of data-availability gives rise to the need for development of empirical research with the aim of valuation approaches to be deployed in the face of this data-scarcity. Responding to this, econometric-techniques demonstrate limitations, as revenue-and-risk-based OLS-techniques demonstrate substantial hidden-variable bias. Meanwhile, as many categories, groups, regions, and clusters known to have explanatory-power have sparsely-available concrete economic-figures which might explain these valuation-differences, fixed-effects demonstrate decreasing marginal explanatory-power as these groups are included and accounted-for. This gives rise to limitations in estimation-accuracy.

To address these methodological-limitations, one could combine known firm-performance indicators and market-conditions such as growth-rates, business cycles, and risk-premiums, with predictive-segmentation of categorical variables, and examination of key fault-lines in the startup landscape.

Such approaches are already used in markets, where an often-used startup-valuation approach are scorecard models, input-based valuation models driven by aggregation of discreet-inputs, and specific discreet contextual-characteristics in which the startups arise. While these approaches traditionally have limited generalizability and have accordingly not made much of an impact in the peer-review landscape, fields such as marketing, psychology, and political science have made extensive use of similar approaches.

This paper's focus is multifold. First, this study brings recent developments in methodology to bear for valuation of startups using machine-learning approaches. Second, this study aims to shed light on divergences between classical valuation-approaches and scorecard and segmented-approaches used in industry. Most importantly, recent developments in machine-learning approaches make possible the hierarchical ranking of valuation-factors, thereby minimizing information-asymmetries, enabling more insightful decision-making. Thirdly, this study seeks to explore the viability of the use of microtargeting on the basis of non-numerical factors such as geographical, business-model-based, and sectoral factors to predict startup-valuations to a high degree of accuracy, overcoming firm-level numerical-factor-reliance.

While it has long been established in both classical economic theory and in entrepreneurial finance that prices and valuations are driven by discreet, measurable factors such as revenues and risk-metrics, the importance and need for hierarchical-ranking as outlined by Quintero (2019), represents an important and emerging gap in the literature, which can be addressed via machine-learning.

This study proceeds as follows: The following section examines the relevant literature, while Section 3 describes the dataset and key variables. Subsequently, Sections 4 and 5 describe the research questions, model-approaches, and findings. Lastly, Section 6 discusses findings and concludes.

## 2. Literature Review

According to Berre and Le Pendeven (2022), the heritage of contextual-based valuation approaches lies with scorecard valuation approaches. Examples of this can be found both in peer-review literature such as Hand (2005) and Sievers et al. (2013), as well as in grey-literature such as Payne (2011) and Berkus (2016), and has made a substantial impact among practitioners in the private equity and venture capital industries, according to Ernst & Young (2020).

Building on this, Quintero (2019) describes the importance and relevance of data-science for the establishment of factor-hierarchy in order to accurately predict valuations within the venture capital industry, specifically suggesting the use of either Bayesian or machine-learning algorithm approaches.

The validity of machine-learning approaches for finance and for entrepreneurial-finance has also been expressed in a growing body of published research focusing on several key topics. This includes use of machine-learning for prediction of both pre- and post-money valuation, for sectoral-clustering, and for successful-exit (Ang et al., 2020). Furthermore, while appropriate- model-selection represents a unique challenge among machine-learning practitioners, Ang et al., (2020) and Quintero (2019) demonstrate that various machine-learning approaches, including Least-Absolute-Shrinkage-and-Selection Operator (LASSO), Latent Dirichlet Allocation (LDA), Categorization-and-Regression-Tree (CART), Extreme-Gradient-Boosting (XGBoost), and Bayesian-approaches such as No-U-Turn Sampler (NUTS) and Markov-Chain-Monte-Carlo (MCMC).

Adding to this, policy sources have also recently begun publishing similar sentiments and views for applications of economy policy ranging from monetary policy to macroprudential policy, to competition policy. Jarmulska (2020) is an ECB study comparing the explanatory-power of random forests to those of more traditional econometric techniques for predicting financial-stress, and finds that the policy-level of decision-tree-based machine-learning approaches are highly-accurate for financial-prediction purposes, even in the face of high-dimensionality macrofinancial and macroeconomic data. Meanwhile, similar techniques are applied to cartel-detection in a competition-policy setting (Huber and Imhof, 2019), for the purposes of competition-law-screening, focusing both competition-enforcement in convoluted settings, and acting as a source of actionable-intelligence and demonstrable-evidence. Here again, accurate actionable-predictions are possible, even in the face of high-dimensionality data.

Mechanically, machine-learning algorithm-reliance on optimal arrangement of multiple decision-factors are central to their functionality. In parallel, multiple-criteria decision analysis (MCDA), which explicitly evaluates conflicting-criteria in decision-making is described Zopounidis et al. (2015) as being used for portfolio and investment-evaluation and selection, is usually implemented in terms of fundamental-factors. While valuation-factors do not necessarily conflict, the valuation-impact of tradeoffs and fault-lines may form important valuation model-elements.

Additionally, several key studies in the economics, finance and entrepreneurial fields bring into context the focus on industry-level, city-level, and business-model focus. Globally, a major debate among the literature is whether the "horse", rather than the "jockey" is more likely to drive investment and valuation. While studies agree that "jockey" refers to management-teams, studies describing the "horse" as industry-level market-conditions, technological-standards, and market-size, (Gompers and Lerner, 2001; Kaplan et al., 2009) outline the horse as being most-important, while Gompers et al. (2020), who describe horse as business-model and firm-related factors, find horse-factors secondary to jockey-factors.

City-level impacts on valuation can be traced to local-level competitive-environments, supply-chain and consumer sophistication, and local factor-conditions ranging from infrastructure to human-capital and workforce-specialization (Porter, 1990).

Meanwhile, Damodaran approaches the valuation landscape with a specifically industry-level outlook. Both Damodaran (2002) and (2009) approach various valuation techniques using sectoral industry-level figures to drive valuations and analysis. While Damodaran (1993) explicitly finds the effect of insiders, Damodaran also publishes aggregated industry-level figures for insider, CEO, and institutional holdings.

For valuation approaches useful to incorporate all of these drivers, Berre and Le Pendeven (2020), a systematic literature review which examines entrepreneurial finance literature and elaborates a multi-step valuation meta-model which describes the mechanical process by which start-up valuation emerges. While the meta-model can accommodate a wide-range of start-up valuation drivers.

## 3. Methodology and Data

**Locally-Supplied Proprietary Data and Commercially-Available Data**

The data consist of proprietary venture-capital deal-data shared by Early Metrics, a Paris-based startup ratings and research agency. To this, we add EU-and EEA-located startup-deals drawn from EIKON and Crunchbase. These supplementary sources were chosen due to their content-similarity to Early Metrics data. To further enrich the dataset, each deal was cross-referenced with firm-performance, industry-level, municipal, and national-level macroeconomic data, and including both proprietary, and commercially-available data, while also boasting extensive variety of value-adding categorical-variables. While the categorical-variables have substantial explanatory-power in their own right, they also add value by interaction with firm-characteristics, as well as macroeconomic and business-cycle market-conditions.

European data grants numerous advantages, such as institutional and macroeconomic diversity meaningful for geographic fixed-effects taking into taking into account of distinct contextual and geographic factors. Because the EEA market is a large developed private equity market where the Common Law, French Civil Law, German Civil Law, and Scandinavian legal families are represented, a European dataset provides the depth of institutional diversity needed to carefully examine the valuation-impact of institutional factors tied to legal origin in a meaningful way. Additionally, as outlined by Berre and Le Pendeven (2020), a research-gap exists concerning model-inclusion of contextual and geographic factors.

Because each line within our dataset is specific per-investor-per-deal, deals with multiple investors occupy multiple lines within the dataset, identifying data for startup and investor, as well as relevant industry-level, institutional, and macroeconomic data for both parties. Since a start-up can have several investors, it can have multiple observations in the regression analysis, reflecting each unique investor–startup pair. The dataset style is borrowed from Masulis and Nahata (2009). With 1,089 observations representing 1,042 deals across 673 startups ranging from Q1-2000 to Q1-2020, our dataset-size is substantial, although only 582 observations contain firm-level revenue figures. Nevertheless, this yields regressions with substantial degrees of freedom compared to prominent studies in the entrepreneurial finance and startup field, such as Gompers et al. (2020), Greenberg (2013), and Masulis and Nahata (2009), who examine 444, 317, and 273 observations respectively. This is outlined in Table 1.

*Table 1: Dataset Observations*

| Source | Observations |
|---|---|
| EIKON | 397 |
| Early Metrics | 80 |
| Crunchbase | 614 |
| **Total** | **1091** |

**Dependent Variables**

The primary dependent variable in this study is pre-money valuation, with valuations expressed in EUR. From EIKON and Early Metrics, the data were collected in Euro, while Crowdcube data was converted from GBP, while Crunchbase data was converted from various currencies for deals in which valuations in EUR were not available. Table 2 outlines the summary statistics of our pre-money valuations data. While the data's time and sectoral distribution is somewhat uneven, it does cover several major events, including the end of the dotcom bubble, the Eurozone crisis, and the start of the Covid-19 Pandemic.

*Table 2: Summary Statistics*

| | 2000 | 2001 | 2002 | 2003 | 2004 | 2005 | 2006 | 2007 | 2008 | 2009 | 2010 | 2011 | 2012 | 2013 | 2014 | 2015 | 2016 | 2017 | 2018 | 2019 | 2020 | Totals |
|---|---|---|---|---|---|---|---|---|---|---|---|---|---|---|---|---|---|---|---|---|---|---|
| Mean Valuation | 109 000 000 | 49 700 000 | 32 900 000 | 21 800 000 | 50 800 000 | 12 000 000 | 25 700 000 | 15 200 000 | 4 181 531 | 511 000 000 | 11 900 000 | 33 300 000 | 40 100 000 | 10 300 000 | 65 400 000 | 182 000 000 | 90 100 000 | 298 000 000 | 286 000 000 | 1 130 000 000 | 1 170 000 000 | 222 000 000 |
| Std. Dev | 239 000 000 | 155 000 000 | 29 800 000 | 15 500 000 | 29 400 000 | 7 679 114 | 32 100 000 | 9 378 568 | 3 352 995 | 690 000 000 | . | 9 424 469 | 135 000 000 | 36 100 000 | 234 000 000 | 363 000 000 | 232 000 000 | 634 000 000 | 787 000 000 | 1 300 000 000 | 1 420 000 000 | 602 000 000 |
| Min | 795 216 | 83 579 | 57 287 | 6 969 987 | 30 000 000 | 762 500 | 2 970 006 | 1 708 426 | 762 500 | 1 558 441 | 11 900 000 | 9 999 972 | 55 714 | 53 833 | 50 000 | 63 330 | 63 330 | 80 000 | 2 160 000 | 1 412 881 | 2 354 802 | 50 000 |
| Max | 1 490 000 000 | 1 280 000 000 | 120 000 000 | 43 700 000 | 71 500 000 | 16 500 000 | 48 400 000 | 24 000 000 | 8 899 903 | 1 270 000 000 | 11 900 000 | 36 700 000 | 556 000 000 | 229 000 000 | 1 560 000 000 | 1 920 000 000 | 1 920 000 000 | 2 140 000 000 | 2 230 000 000 | 5 380 000 000 | 5 060 000 000 | 5 380 000 000 |
| Austria | - | 3 | - | - | - | - | - | - | 2 | - | - | - | - | - | 1 | - | - | - | - | - | - | 6 |
| Belgium | - | 3 | - | - | - | - | - | - | - | - | - | - | - | - | - | 1 | - | 3 | 5 | - | 5 | 17 |
| Croatia | - | - | - | - | - | - | - | - | - | - | - | - | - | - | - | - | - | - | - | 1 | - | 1 |
| Czech | 6 | - | - | - | - | - | - | - | - | - | - | - | - | - | - | - | - | - | - | - | - | 6 |
| Denmark | - | 3 | - | - | - | - | - | - | - | - | - | - | - | 1 | - | - | - | 1 | - | 3 | - | 8 |
| Finland | 1 | - | - | - | - | - | - | - | - | - | - | 7 | - | 1 | 1 | 5 | 2 | 4 | 1 | - | - | 22 |
| France | 29 | 11 | 4 | - | - | 5 | - | - | - | 2 | - | 1 | 1 | 4 | - | 5 | 27 | 16 | - | 1 | - | 106 |
| Germany | 19 | 11 | 3 | - | 1 | - | - | - | 2 | - | - | - | 2 | 3 | 5 | 12 | 14 | 33 | - | 31 | 2 | 138 |
| Ireland | - | 1 | - | - | - | - | - | - | - | - | - | - | - | - | 1 | 5 | 8 | 5 | - | - | - | 20 |
| Italy | 3 | - | - | - | 1 | - | - | - | - | - | - | - | 1 | 1 | - | 1 | 2 | 1 | - | - | - | 10 |
| Latvia | - | - | - | - | - | - | - | - | - | - | - | - | - | - | - | 1 | - | - | - | - | - | 1 |
| Lithuania | - | - | - | - | - | - | - | - | - | - | - | - | - | - | - | - | - | - | - | 7 | - | 7 |
| Luxembourg | 1 | - | - | - | - | - | - | - | - | - | - | - | - | - | - | 2 | 3 | - | - | - | - | 6 |
| Netherlands | 2 | 1 | 2 | - | - | - | - | - | - | - | - | - | - | - | 5 | - | - | 1 | - | - | - | 11 |
| Norway | 2 | 4 | - | - | - | - | - | 2 | - | 1 | - | - | - | - | - | - | - | - | 1 | - | - | 10 |
| Poland | 9 | 8 | - | - | - | - | - | - | - | - | - | 1 | - | - | 1 | 1 | 2 | - | - | - | - | 22 |
| Portugal | - | - | - | - | - | - | 1 | 1 | - | - | - | - | - | - | - | 1 | 2 | 2 | 1 | - | - | 8 |
| Romania | 4 | - | - | - | - | - | - | - | - | - | - | - | - | - | - | - | - | - | 2 | - | - | 6 |
| Spain | 5 | - | 1 | - | - | - | - | - | - | - | - | - | - | 2 | 1 | 2 | 4 | 2 | - | 4 | - | 21 |
| Sweden | 5 | 3 | - | - | - | - | - | - | - | - | - | 1 | 1 | - | 1 | 13 | - | 2 | - | 1 | - | 27 |
| Switzerland | 4 | 1 | 1 | - | - | - | - | - | - | - | 1 | - | - | - | 1 | - | 2 | 3 | - | - | - | 13 |
| UK | 43 | 59 | 6 | 4 | - | 2 | 1 | 2 | 1 | 1 | - | - | 12 | 30 | 89 | 151 | 116 | 54 | 5 | 26 | 21 | 623 |
| Total | 133 | 108 | 17 | 4 | 2 | 7 | 2 | 5 | 5 | 5 | 1 | 9 | 17 | 42 | 106 | 200 | 182 | 130 | 12 | 73 | 28 | 1089 |
| Business / Consumer Services | - | - | - | - | - | - | - | 1 | - | - | - | - | 5 | 8 | 30 | 33 | 21 | 12 | - | - | - | 110 |
| Aerospace | - | - | - | - | - | - | - | - | - | - | - | - | - | - | - | - | - | 1 | - | - | - | 1 |
| Retail | 12 | 9 | - | - | - | - | - | - | - | - | - | - | 1 | 3 | 9 | 22 | 31 | 2 | - | 7 | - | 96 |
| Automotive | - | 2 | - | - | - | - | - | - | - | - | - | - | 1 | 1 | - | 2 | 9 | - | - | - | - | 15 |
| Finance | 5 | 4 | 2 | - | 1 | - | - | 1 | - | 1 | - | - | 1 | 4 | 5 | 45 | 23 | 23 | 1 | 17 | 4 | 137 |
| Food/Agro | 3 | - | - | - | - | - | - | - | - | - | - | - | - | 4 | 16 | 12 | 16 | 16 | 1 | - | 1 | 69 |
| Machinery / Industrial | 6 | 4 | 1 | 2 | - | - | - | - | - | - | - | - | - | 1 | - | 2 | 5 | 3 | - | - | - | 24 |
| Power | 1 | 1 | - | - | - | - | - | - | - | - | - | - | 4 | - | - | - | - | - | - | - | - | 6 |
| ICT / Software | 62 | 45 | 6 | 1 | - | 5 | 2 | 2 | 3 | - | - | 9 | 5 | 9 | 23 | 27 | 35 | 33 | 5 | 36 | 18 | 326 |
| Pharma / Healthcare | 18 | 21 | 7 | - | 1 | 2 | - | - | 2 | 1 | 1 | - | - | 1 | - | 11 | 13 | 3 | 1 | 1 | 5 | 88 |
| Education | - | 1 | - | - | - | - | - | - | - | - | - | - | 1 | - | 1 | 3 | 4 | 3 | 1 | - | - | 14 |
| Electronics | 2 | 1 | - | - | - | - | - | - | - | - | - | - | 1 | 2 | - | 3 | 2 | 1 | 1 | - | - | 13 |
| Leisure / Entertain/ Tourism | 5 | 3 | 1 | - | - | - | - | - | - | - | - | - | 2 | 2 | 10 | 15 | 18 | 5 | - | 3 | - | 64 |
| Clean-Tech | - | 1 | - | - | - | - | - | 1 | - | - | - | - | 1 | - | - | 1 | 1 | 4 | 2 | - | - | 11 |
| Home | - | - | - | - | - | - | - | - | 1 | - | - | - | - | - | - | 1 | - | - | - | - | - | 2 |
| Real Estate | 1 | 2 | - | - | - | - | - | - | - | - | - | - | - | - | 1 | 11 | 3 | 8 | - | 9 | - | 35 |
| Office | - | - | - | - | - | - | - | - | - | - | - | - | - | - | - | - | 2 | - | 1 | - | - | 3 |
| Fossil | - | 1 | - | 1 | - | - | - | - | - | - | - | - | - | - | - | - | - | - | - | - | - | 2 |
| Transport | 1 | - | - | - | - | - | - | - | - | 2 | - | - | - | 2 | 1 | 11 | 6 | 6 | - | - | - | 29 |
| Media | 7 | 3 | - | - | - | - | - | - | - | - | - | - | - | - | 1 | 8 | 1 | - | - | - | - | 20 |
| Total | 123 | 98 | 17 | 4 | 2 | 7 | 2 | 5 | 5 | 5 | 1 | 9 | 17 | 42 | 105 | 200 | 180 | 130 | 12 | 73 | 28 | |

**Independent Variables**

In principle, independent variables are available as either continuous variables such as revenues, and credit-risk premia, or as categorical-variables such as industry, investor-type, country or city. Because the valuation-impact of continuous variables on startups is both widely-documented and easy to measure, this study's focus will be on the valuation-impact of categorical-variables.

**Independent Variables: Firm Characteristics**

*Categorical Variables*

Geographical information variables have gathered substantial attention in the press covering the startup landscape and the venture capital industry. To that effect, the dataset contains two levels of granularity, cities, and countries. In principle, this is able to capture the valuation impacts of Europe's startup hotspots such and London, Paris, Berlin, Stockholm, and Helsinki, while also keeping track of country-level startup-markets, such as those in the Benelux countries and Scandinavian countries.

Beyond that, the dataset also contains information for sectoral industry. While it is widely-reported that industries such as biotech, fintech, and renewables attract both public and investor attention, the valuation impact is not totally explained by variations in revenue.

*Continuous Variables*

Revenue figures used for this study are cross-sectional. Limitations in publicly-available data gave rise to the need to rely on cross-sectional revenue figures, rather than time-series revenue figures. These figures were gathered from EIKON, Zoominfo, and Dun & Bradstreet. While EIKON figures were collected in EUR, all other figures were converted into EUR using exchange rates appropriate for the year in which the deal occurred. Intellectual property meanwhile, was collected from the European Patent Organization's EPO PATSTAT database. It consists of the number of patent documents recognizing the start-up as the patent holder.

**Independent Variables: Investor Characteristics**

*Categorical Variable*

As is the case with firm characteristics, geographic information is available at two levels of granularity, capturing cities, and countries, since it might be the case that investor-location may matter more to startup-valuation than startup-location does. Or that geographic interactive-effects might play a role. Beyond that, investor-type is also available within the dataset, capturing both GVCs and CVCs, as well as their likely impacts on valuations.

**Independent Variables: Market Conditions**

The market conditions available in our database is extensive. These consist of national-level data, which is both macroeconomic and institutional, as well as municipal-level data, and sectoral industry-level data.

The macroeconomic data consist of output gap and total venture capital financing at county level. These are drawn from the OECD and IMF. Alternate estimate figures for and total venture capital financing at county level are synthesized by aggregating all deals country-level deals data from Crunchbase for each country in the dataset.

Institutional data are drawn primarily from two sources. La Porta et al, provide legal protection indices for self-dealing and investor protection. The World Economic Forum's World Competitiveness Report, which not only yields a competitiveness index, but also indices for macroeconomic stability, trade-openness, and availability of SME funding. Use of European deals data allow us to draw on a wide-range of variation in these indicators.

At a more granular level, city-level data also exist within the dataset. Per-capital income levels may help to explain local-level consumer market-size, which can drive both startup revenues and investor sizes. Meanwhile, infrastructure measures may serve to attract both investors and entrepreneurs to a city. Fundamentally, Porter (1990) describes both as critical to the formation of economic clusters.

## 4. Hypothesis and Analysis:

At a general level, startup-valuation is known to be driven primarily by measurable determinants such as revenues which can be discounted, risk-measures which impact risk-adjusted discount-factors (Damodaran 2009), and assets (Damodaran 2010), which drive output and productivity, while also being tradable to markets and investors.

Nevertheless, market-conditions are also understood to play an important role in determining VC investment (Gompers et al. 2020), with industry, business model, and even economic geography being almost as important to investors as firm performance and entrepreneurial team. On the other hand, valuations can be also driven by market-conditions more-specific to industry-level sectoral conditions (Damodaran, 2009), as well as local-level conditions relating to local-area specialization, competitive-conditions, market-sophistication, and entrepreneurial ecosystem (Porter, 1990).

Overall, value can be detected by, and with respect to contextually-relevant categorical variables. This means that in absence of revenue data, contextual information can be used to construct a scorecard approach to valuation. It may be the case that these variables rive the valuation, or that they simply serve as indicators.

While several sources of divergence in firm-performance, strategy, and valuation have been enumerated by the literature, startup-valuation literature focuses particularly on sectoral and industry-level data. Damodaran (2009) for example, takes into consideration industry-level risk-premiums, survival rates, and governance variables. For European startup-investment deals outlined in this study's dataset, this can be tested by measuring valuation-impact of industry-level valuation-drivers, and examining both the causal valuation-impacts, as well as the key fault-lines, sub-populations, and order of variable-importance via both OLS and CART approaches. These can then be compared in terms of explanatory power, fault-lines and variable-importance-order to industry-level microtargeting using categorical-variables to capture and describe similar populations of startups and deals. This leads us to:

## H1: Industry tells the story

*Key insights can be gathered by examining industry-level categoricals. Key fault-lines exist, which play a deterministic role on valuations, and which are unexplained by discounted-cashflow (DCF) and by continuous industry-level variables.*

Beyond reliance on sectoral industry-level indicators, economic-geography is also a major focus of both startup-literature, such as Nathan and Vandore (2014) and startup-related press, such as TechCrunch and Business Insider, who focus on the current-events of startup markets. Mechanically-speaking, Porter (1990) examines in detail the architecture of local-level economic-geography, while national-level economic-geography, which might be driven by business-cycles (Heughebaert and Manigart, 2012; Streletzki and Schulte, 2013), venture-capital dynamics and market-size (Inderst and Muller, 2004), or macrofinancial conditions (Bonini and Alkan, 2009), is examined by multiple authors. In principle, the role of economic-geography can be examined by measuring the valuation-impact of local-level and country-level valuation-drivers, and examining both the causal valuation-impacts, as well as the key fault-lines, sub-populations, and order of variable-importance via both OLS and CART approaches. These can then be compared in terms of explanatory power, fault-lines and variable-importance-order to geographic microtargeting using categorical-variables to capture and describe similar populations of startups and deals. This leads us to:

## H2: Countries and Cities tell the story

*Key insights can be gathered by examining the available local and national-level categorical variables. Key fault-lines exist, which play a deterministic role on valuations, and which are unexplained by discounted-cashflow (DCF) and by continuous industry-level variables.*
Because Fama (1970), describes that market-prices reflect *all* valuation-factors, which can take the form of any discreet information-package, as well as risks, uncertainties, and ambiguities, it stands to presume that all valuation-factors ranging from country-level indicators to sector and local-level indicators can act in unison alongside firm-level variables to grant insight into startup-valuations. In principle, these impacts can be traced via use of multi-factor regression models, which aim at factor-saturation by including cross-sections of sectoral, regional, and national-level data in addition to firm-level indicators. Nevertheless, a multi-factor approach is limited in its ability to use joint-categorical variables. Fixed-effects regressions require substantial observation numbers on a per-category basis in order to deliver accurate insights, giving rise to limitations in estimation-accuracy otherwise, flowing from small group-sizes, which may yield extensive Type-1 errors, particularly when not offset by a very large number of groups (Wooldridge, 2010; Theall et al., 2011). Mechanically, this limitation can be overcome via the use of CART-based approaches for combined categorical variables, leading us to:

## H3: It takes a combination of categoricals to convey an accurate valuation

*It takes a combination of categorical valuation-determinants to convey an accurate valuation: We rely on joint-variables and interaction effects for key insights.*

# 5. Model Approach and Empirical Findings:

Established economic theory outlines that valuation acts as a price-signal that communicates revenues, risks, and firm key fundamentals. However, the startup landscape is generally characterized by the difficult and sparse availability of accurate and stable revenue and risk data, which valuations might otherwise signal to markets. This reality has led to the rise of a diverse taxonomy of models, views and valuation approaches both in peer-review literature and among private equity, VC and business angel investors.

## Model Approaches to Valuation

### Discounted Cashflow Approach

The discounted-cashflow (DCF) approach is among the most traditionally established valuation approaches for both firms and individual assets. Damodaran (2009) outlines that while application is not as straightforward in the context of start-ups as is the case with more-established firms, which can often include unstable revenue-estimates, risk-metrics, and growth-rates estimates, potentially yielding unrealistic valuation estimates.

### Relative Valuation Approach

The relative valuation approach, which measures firm valuation relative to a either firm's performance indicators or balance-sheet figures, estimating firm-valuation of an asset by examining pricing of comparable assets relative to a common variable such as earnings, assets, revenues, cashflows, or book value. Damodaran (2002) outlines that the relative-valuation approach assumes that industry-peers and comparable firms to the firm being valued are in general, being priced correctly by the market.

### How Contextual Factors Play a Role

A useful theoretical approach used by a minority of scholars is that of the scorecard-based approach. The primary advantage of the scorecard approach is the ability to incorporate qualitative, geographic, sectoral or categorical determinants in several ways ranging from the non-financial and deal-characteristics prevalent in a given sectoral or municipal ecosystem, to the role of national-level or market-condition. This approach is capable of shedding light into valuation even as detailed related economic and financial information is missing, scare, or unevenly available.

Scorecard-based valuation-methods are modular and relatively straightforward valuation-approaches based on summation of key characteristics, market-conditions, and deal-conditions developed mainly by industry practitioners. In industry, the scorecard approach is typically used by business angels. Industry-emergent techniques for scorecard valuation include Berkus (2016) and Payne (2011).

Meanwhile, in published economic literature, this same concept appears as summation-based valuation models, such as published by Hand (2005) and Sievers et al. (2013). For example, Eq. 1 outlines the Sievers et al. (2013) summation-based valuation model, assigning valuation based on summation of financial, and non-financial firm-attributes, as well as deal-characteristics are relevant valuation-coefficients.

*Equation 1: Sievers et al. (2013) Summation-based Valuation Model*

$$\log(Valuation_{it}) = \sum \Phi Non-financial_{it} + \sum \Delta Financial_{it} + \sum \Psi Deal\ Characteristics_{it}$$

The recent emergence of machine-learning techniques has led to increasing methodological sophistication of scorecard approaches, as predictive-techniques incorporating categorical, geo-spatial, and qualitative data become widespread. Microtargeting by means of data-mining, for example can allow scorecard-based valuation-approaches to incorporate qualitative and categorical data hierarchically, and to a potentially-extreme degree of detail.

Functionally speaking, a hierarchically-structured valuation-model that would result from a microtargeting approach can be elaborated as staged-valuation approach, such as the Startup-Valuation Meta-Model developed by Berre and Le Pendeven (2022).

Formally:

*Equation 2: Berre and Le Pendeven (2022) Startup Valuation Meta-Model*

$$Pre-Money\ Valuation = f(((\sum Start-Up\ Value) \sum Deal\ Value) \sum Deal\ Valuation)$$

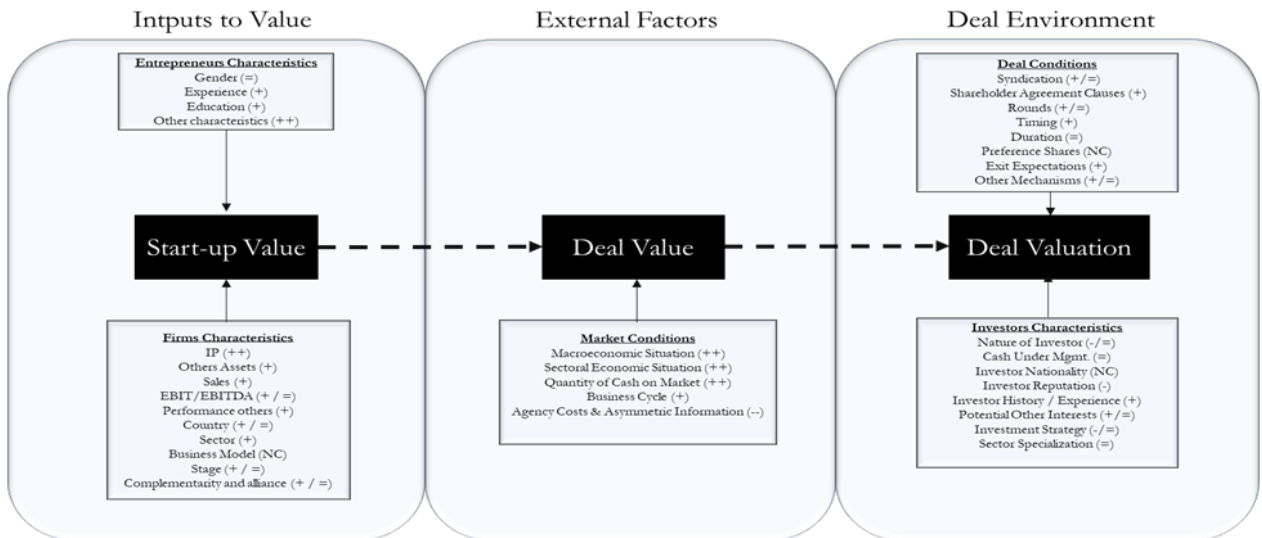

A key advantage of this type of approach, is that outputs such as valuation, selection, or survival probability can be microtargeted by including ever smaller and more specific categorical information, or combinations thereof.

**Microtargeting: An Alternate Approach**

Mechanically, microtargeting by means of data-mining is described by Murray and Scime (2010), as the process of inductively analyzing data to find actionable-patterns, fault-lines and relationships within the data, on the basis of trends related to both numerical-characteristics, such as number of family members, average family age, and descriptive-characteristics such as geographic-area, via construction of decision trees, which is an analytical-technique that is both explanatory and predictive, and is used for both variable-predictions, as well as to provide knowledge about structure, segmentation, and interrelationships among data. This approach provide insight into how the outcome variable's value is dependent on the deterministic-factors, with each identifiable fault-line constituting a segment of individuals. According to Murray and Scime (2010), these characteristics have given microtargeting approaches a long history in both marketing and political-science.

**Comparing and Combining Approaches**

Overall, the accuracy of regression-tree models can be compared to those of equivalently-constructed regression models on the basis of their goodness-of-fit indicators. While linear-regressions are typically evaluated on the basis of $R^2$, Berre (2023) describes that regression-trees can be evaluated on the basis of $1 - R^2$ root-mean-squared-error.

For the purposes of startup valuation, the informational content of descriptive and categorical characteristics such as industry, geography, and business model are often overlooked, despite the general possibility that these characteristics might possess explanatory power equivalent to multiple associated numerical variables. Meanwhile, use of fixed-effects to incorporate descriptive categorical characteristics suffers losses in explanatory power as the number of descriptive characteristics increases, whereas microtargeting approaches improve their accuracy as the number of these characteristics increases.

In contrast, this study includes extensive categorical variables capturing country, city, industry, business model, investor, and ownership, as well as numerical variables attached to country, city, industry, outlined in Table 3.

*Table 3: Variable Definitions*

| Variable | Description |
|---|---|
| Valuation | Pre-Money Startup-Valuation. Source: EIKON, Early Metrics, Crunchbase |
| Revenue | Startup company revenue: Source: Eikon, Dun & Bradstreet, Zoominfo |
| Beta | Unlevered sectoral beta. Source: NYU Stern dataset |
| Credit Risk Premium | Country-Level Credit Risk Premium. Source: Moody's, NYU Stern |
| Output Gap | Deviations of actual GDP from potential GDP as % of potential GDP. Source: OECD |
| Cash-on-Market | Country-level total venture capital investments. Source: OECD |
| Sector Categorical Variable | Categorical Variable. Industry sector of startup. |
| City Categorical Variable | Categorical Variable. City where startup is based. |
| Investor-type Categorical Variable | Categorical Variable. Types of investors involved in deal. |
| Customer-type Categorical Variable | Categorical Variable. Types of customers targeted by the started. |
| Ownership-status Categorical Variable | Categorical Variable. Whether a firm is publicly-traded, privately-held, or a subsidiary |
| City Physical Infrastructure | City-level variable: Measure of how developed the city's infrastructure is. Source Economist Intelligence Unit |
| City Labor Productivity | City-level variable: Measure of City's Labor Productivity. Source: OECD |
| City Intellectual Property | City-level variable: Number of patents filed in the city. Source: European Patents Office |
| No. of deals in the city | City-level variable: Number of deals in the city. Source: Author |
| Municipal GDP | City-level variable: Measure of City's Economy. Source: OECD |
| Industry-Level Insider Holdings | Industry-level variable: Sectoral-level percent of firms held by insiders. Source: NYU Stern |

Additionally, the hierarchical regression-tree approach can also be combined with OLS and fixed-effects regression approaches, as an ex-post corroborational approach, which has the potential to not only corroborate regression-findings, by corroborating both causality and functional-form, but also the potential to elaborate on regression-findings by adding hierarchical factor-determinant insights. So relevant is this elaboration potential that it has the potential to provide insights and actionable-findings, even in the face near-significant or non-significant regression-results, meaning that it can be possible to reach decisive findings in a wider diversity of circumstances using a combined two-tiered approach.

In some circumstances, tree-based corroboration can be reinforced by use of Random Forest to reinforce the regression-tree approach making use of the law of large-numbers to minimize the impact of outliers. This approach would be highly-useful however, when dealing primarily with continuous or discreet numerical variables. The random-forest approach however, has its methodological limitations, since the random random-forest approach has a hard-limit on the number of categories a categorical explanatory variable can have. While this limit is large-enough to accommodate compact categorical factors with relatively-few categories, categorical factors relating to industry or economic-geography can grow to be both category-dense and information-rich. In this dataset, the most consequential and deterministic categorical valuation-factors are too category-dense to be suitable for the random-forest approach. In these cases, an alternate way to mitigate outlier-impact is via log-transformation, which "flattens" by restraining effects of outliers on dataset means and medians. Because regression trees and partitioning methods in general are sensitive to the presence of outlier-influence in terms of dependent-variable outliers (Khan et al., 2013), the flattening of outliers has the potential to add substantial explanatory-power to regression-tree models, as log-transformation reduces estimation-problems associated with percentage changes from baseline (Keene, 1995), while maximizing data-scale-flattening (Ribeiro-Oliveira et al., 2018). Variables showing skewed distribution can also be made symmetric using log-transformation (Keene, 1995).

**Discounted-Cashflow Based Models**

Table 4 outlines valuation-impact of revenues, sector unlevered-beta, and country-risk-premium and demonstrates that the data are consistent with discounted cash-flow valuation models, and thereby establishes the soundness of the dataset. In principle, a DCF approach can be approximated using OLS models by including revenue or profit figures and discount-factor inputs. Because the free-cashflow-to-equity DCF model applies a cost-of-equity-based discount factor, an FCFE-based DCF approach would include CAPM-related factors in the estimation model. Overall, all DCF-factors boast both individual and joint statistical significance. Hereafter, these factors serve as a benchmark against which to compare other models. Overall, this table demonstrates that DCF-factors are both individually and jointly-significant, with goodness-of-fit indicators for the single-factor regressions indicating that revenue has the strongest explanatory-power, followed by sectoral-beta, while the multivariate regressions indicating that roughly 40% of valuation-dynamics are captured by the DCF approach.

*Table 4: Baseline Discounted-Cashflow Approximation OLS*

**DCF-based Regressions**

| VARIABLES | (1) Ln_Valuation | (2) Ln_Valuation | (3) Ln_Valuation | (4) Ln_Valuation | (5) Ln_Valuation | (6) Ln_Valuation |
|---|---|---|---|---|---|---|
| Ln_Revenue | 0.6861*** | | | 0.6660*** | 0.6692*** | 0.6514*** |
| | [0.034] | | | [0.034] | [0.034] | [0.034] |
| Ln_Beta | | -2.7738*** | | -1.1629*** | | -1.0886*** |
| | | [0.395] | | [0.389] | | [0.388] |
| Country Risk Premium | | | -70.7055*** | | -40.7742*** | -38.3685*** |
| | | | [11.701] | | [13.134] | [13.092] |
| Constant | 6.4565*** | 17.8453*** | 16.7791*** | 7.4278*** | 7.0425*** | 7.9171*** |
| | [0.503] | [0.254] | [0.133] | [0.596] | [0.534] | [0.615] |
| Observations | 646 | 1,045 | 1,045 | 646 | 646 | 646 |
| R-squared | 0.39 | 0.05 | 0.03 | 0.40 | 0.40 | 0.41 |
| Adjusted R-squared | 0.393 | 0.0442 | 0.0329 | 0.401 | 0.401 | 0.408 |

Standard errors in brackets
*** $p<0.01$, ** $p<0.05$, * $p<0.1$

Corroborating the OLS and fixed-effects findings outlined in Table 4, Table 5 examines impact of DCF-factors using a CART hierarchical regression tree-approach, agreeing that revenue holds the strongest explanatory-power, followed by sectoral-beta. Compared to the OLS approach outlined in Table 4 however, the CART approach finds that country-risk-premium plays a more deterministic-role in low-revenue and low-valuation startups, whereas sectoral-beta plays a more deterministic-role in high-revenue and high-valuation startups. Root-mean squared-error (RMSE), the CART-approach's the goodness-of-fit indicator demonstrates that Table 5's CART model, which mirrors the sixth regression in Table 6 features substantially-stronger goodness-of-fit, capturing roughly half of the valuation-variation within the dataset.

*Table 5: Baseline Discounted Cashflow Approximation CART*

| OBS: | 1045 | | | |
|---|---|---|---|---|
| End Nodes: | 10 | | | |
| Complexity parameter | No. of Split | RMSE | Crossvalidation error | Crossvalidation St. Dev. |
| 0.31393 | 0 | 1.0000 | 1.00094 | 0.03164 |
| 0.03812 | 1 | 0.6861 | 0.71385 | 0.03123 |
| 0.02337 | 2 | 0.6480 | 0.67459 | 0.02986 |
| 0.02155 | 3 | 0.6246 | 0.66970 | 0.02950 |
| 0.01876 | 4 | 0.6030 | 0.65757 | 0.02940 |
| 0.01534 | 5 | 0.5843 | 0.64304 | 0.02920 |
| 0.01450 | 6 | 0.5689 | 0.61971 | 0.02839 |
| 0.01152 | 8 | 0.5399 | 0.60189 | 0.02763 |
| 0.01152 | 9 | 0.5284 | 0.59218 | 0.02722 |
| 0.01000 | 11 | 0.5053 | 0.58951 | 0.02802 |
| Variable Importance | | | | |
| Ln_Revenue | Ln_Beta | Country-Risk Premium | | |
| 59 | 23 | 18 | | |

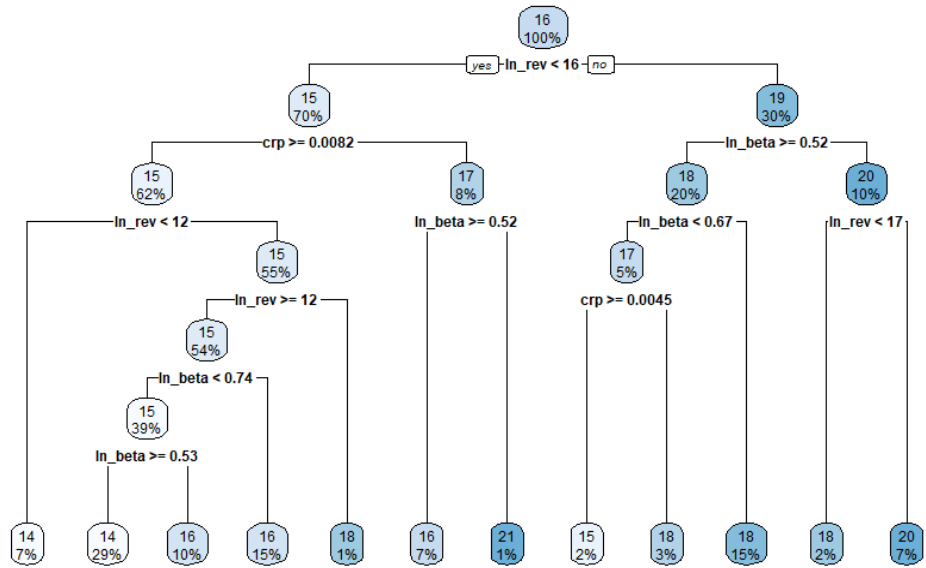

Because all factors in this model are numerical variables, it is possible to build on this compound approach further. Table 6 examines the same model using a random-forest algorithm, and corroborating the relative-variable-importance results reached by the CART approach in Table 5, but increasing goodness-of-fit even further, above 70%.

*Table 6: Baseline Discounted Cashflow Approximation Random Forest*

| Type of random forest : | Regression |
|---|---|
| Number of trees : | 1000 |
| No. of variables tried at each split: | 1 |
| Mean of squared residuals : | 2.268259 |
| % Var explained : | 73.99 |

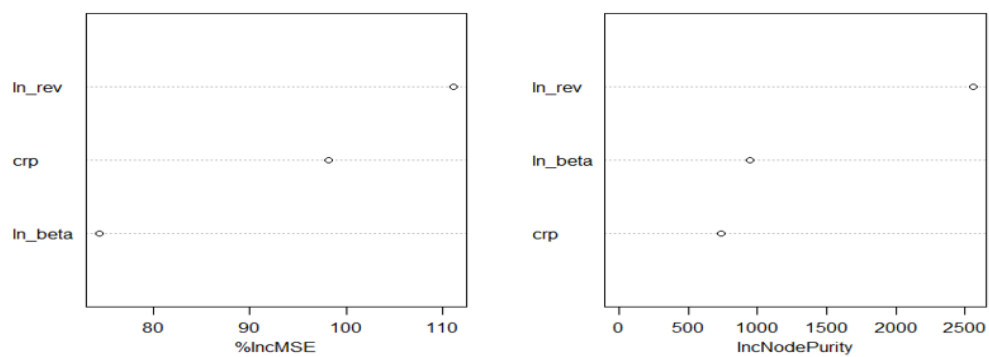

Random Forest Micro-Targeting: DCF Factors

## The Industry Level

Moving beyond DCF-factors, Table 7 examining industry-level variables grants the opportunity to measure valuation-impact of both industry-level insider-holdings and sector-fixed-effects, finding that when sector-level factors are taken into account, startup-valuations are less sensitive to sector-level beta. Sector-fixed-effects outlined in Table 7's third regression also find sector-level beta non-significant, with some loss of overall explanatory-power compared to regressions 1 and 2. This indicates that while sector-level beta has some explanatory-power as per DCF-model, it is overshadowed by both industry-level insider-holdings and by sector-dummies.

*Table 7: Sector-level OLS and Fixed-effects*

**Sector-Level Regressions**

| VARIABLES | (1) Ln_Valuation | (2) Ln_Valuation | (3) Ln_Valuation |
|---|---|---|---|
| Ln_Revenue | 0.6514*** | 0.6694*** | 0.6318*** |
|  | [0.034] | [0.034] | [0.034] |
| Ln_Beta | -1.0886*** | -0.5556 | -0.3805 |
|  | [0.388] | [0.405] | [0.578] |
| Country Risk Premium | -38.3685*** | -36.9577*** | -38.3666*** |
|  | [13.092] | [12.945] | [12.737] |
| Insiders |  | -8.7086*** |  |
|  |  | [2.159] |  |
| Constant | 7.9171*** | 8.6577*** | 7.6960*** |
|  | [0.615] | [0.634] | [0.690] |
|  |  |  |  |
| Observations | 646 | 645 | 643 |
| R-squared | 0.41 | 0.42 |  |
| Adjusted R-squared | 0.408 | 0.420 |  |
| Number of Sectors |  |  | 18 |
| Within-R-squared |  |  | 0.356 |
| Between-R-squared |  |  | 0.691 |
| Overall-R-squared |  |  | 0.409 |

Standard errors in brackets
*** p<0.01, ** p<0.05, * p<0.1

Corroborating the OLS and fixed-effects findings outlined in Table 7, Table 8 examines the impact of DCF factors and both industry-level variables and industry dummies. Overall, Table 8 demonstrates that while DCF factors play a key deterministic role, with firm-revenue being the most influential determinant. Meanwhile the industry-level insider-holdings play a much smaller deterministic role, while the industry categorical variable does not play a significant role. Meanwhile, RMSE indicates roughly-equal explanatory power compared to the OLS model outlined in Table 7, as well as the within-sector goodness-of-fit indicator, but not the between-sector goodness-of-fit indicator of the fixed-effects regression.

*Table 8: Sector-focused CART*

| OBS: | 1045 | | | |
|---|---|---|---|---|
| End Nodes: | 11 | | | |
| Complexity parameter | No. of Split | RMSE | Crossvalidation error | Crossvalidation St. Dev. |
| 0.31393 | 0 | 1.0000 | 1.00050 | 0.03165 |
| 0.03812 | 1 | 0.6861 | 0.69393 | 0.02939 |
| 0.02337 | 2 | 0.6480 | 0.65629 | 0.02762 |
| 0.02155 | 3 | 0.6246 | 0.65061 | 0.02753 |
| 0.01876 | 4 | 0.6030 | 0.62361 | 0.02650 |
| 0.01539 | 5 | 0.5843 | 0.61061 | 0.02606 |
| 0.01534 | 6 | 0.5689 | 0.60005 | 0.02623 |
| 0.01332 | 7 | 0.5535 | 0.59658 | 0.02619 |
| 0.01152 | 9 | 0.5269 | 0.59167 | 0.02680 |
| 0.01000 | 10 | 0.5154 | 0.57010 | 0.02565 |
| Variable Importance | | | | |
| Ln_Revenue | Ln_Beta | Country-Risk Premium | Insider Holdings | |
| 54 | 17 | 16 | 12 | |

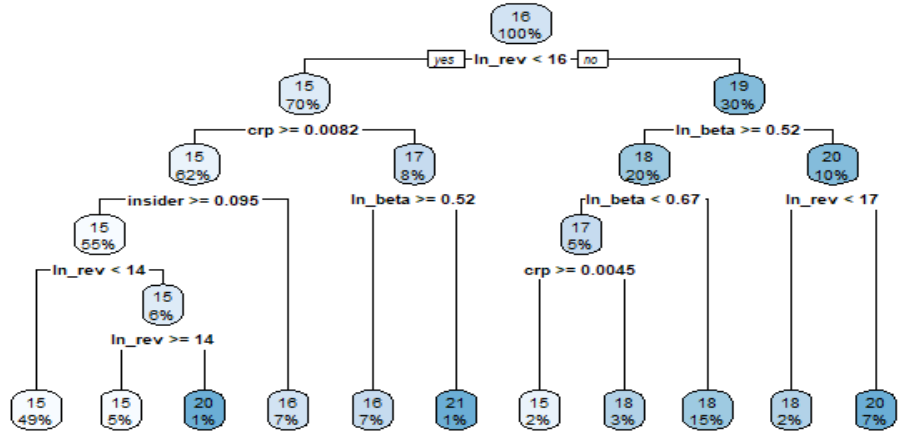

On the other hand, the microtargeting-based CART approach outlined in Table 9, replacing industry-level numerical-variables (aside from sectoral-beta, a DCF-factor) with industry categorical-variables yields virtually-identical model goodness-of-fit, but finds somewhat different variable-importance hierarchy, with sector not only having higher variable-importance, but also playing among high-revenue, high-valuation startups. While individual sector-level variables play relatively-minor deterministic roles, the sector categorical-variable has substantially more explanatory-power. This indicates that the categorical sector-variable has substantially more information-content than either industry-level beta or industry-level insider holdings.

*Table 9: Sector-focused CART-based microtargeting*

| OBS: | 1045 | | | |
|---|---|---|---|---|
| End Nodes: | 10 | | | |
| Complexity parameter | No. of Split | RMSE | Crossvalidation error | Crossvalidation St. Dev. |
| 0.31393 | 0 | 1.0000 | 1.00145 | 0.03167 |
| 0.04611 | 1 | 0.6861 | 0.70314 | 0.03007 |
| 0.04011 | 2 | 0.6400 | 0.67444 | 0.03011 |
| 0.02123 | 3 | 0.5999 | 0.64082 | 0.02908 |
| 0.02093 | 4 | 0.5786 | 0.63841 | 0.02871 |
| 0.01408 | 5 | 0.5577 | 0.62109 | 0.02808 |
| 0.01283 | 7 | 0.5295 | 0.58447 | 0.02726 |
| 0.01162 | 8 | 0.5167 | 0.56915 | 0.02706 |
| 0.01000 | 9 | 0.5051 | 0.56886 | 0.02721 |
| Variable Importance | | | | |
| Ln_Revenue | Sector | Ln_Beta | Country-Risk Premium | |
| 54 | 17 | 16 | 12 | |

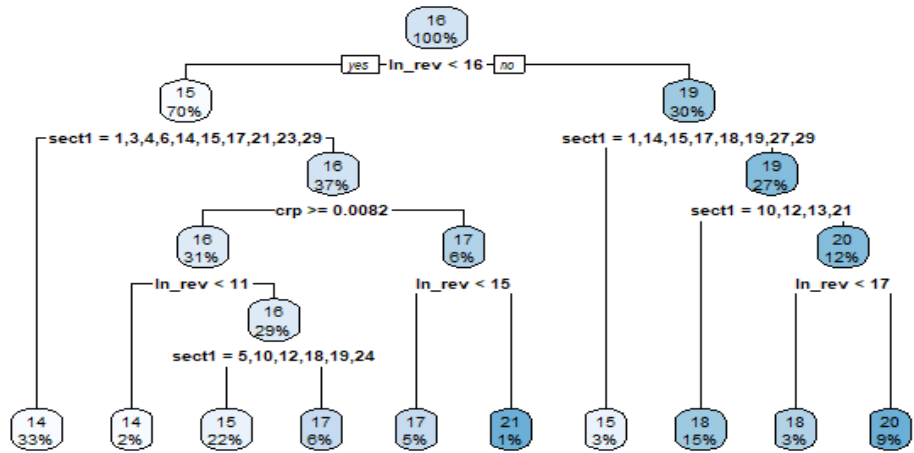

## The Country Level

While the DCF model does include country-level determinants to influence the valuation's discount rate, by including country-risk premiums, several studies also describe the valuation-impact had by additional national-level variables such as business cycle indicators, or VC market size. Table 10 outlines what OLS and country-fixed-effects regressions can demonstrate concerning the valuation-impact of adding country-level valuation determinants.

*Table 10: Country-level OLS and Fixed Effects*

| Country-level Regressions | | | | | |
|---|---|---|---|---|---|
| | (1) | (2) | (3) | (5) | (4) |
| VARIABLES | Ln_Valuation | Ln_Valuation | Ln_Valuation | Ln_Valuation | Ln_Valuation |
| Ln_Revenue | 0.6514*** | 0.6862*** | 0.6512*** | 0.6792*** | 0.5817*** |
| | [0.034] | [0.036] | [0.035] | [0.036] | [0.034] |
| Ln_Beta | -1.0886*** | -0.9448** | -1.1753*** | -1.0286*** | -1.5274*** |
| | [0.388] | [0.390] | [0.394] | [0.393] | [0.379] |
| Country Risk Premium | -38.3685*** | -23.8422* | -13.3416 | -11.4942 | -6.3079 |
| | [13.092] | [14.386] | [15.048] | [16.327] | [21.354] |
| Cash on Market | | 0.0006*** | | 0.0006*** | |
| | | [0.000] | | [0.000] | |
| Output Gap | | | 0.2101*** | 0.1216 | |
| | | | [0.070] | [0.075] | |
| Constant | 7.9171*** | 6.4131*** | 7.9414*** | 6.6156*** | 9.3378*** |
| | [0.615] | [0.648] | [0.634] | [0.659] | [0.691] |
| Observations | 646 | 578 | 632 | 577 | 646 |
| R-squared | 0.41 | 0.48 | 0.42 | 0.48 | |
| Adjusted R-squared | 0.408 | 0.473 | 0.421 | 0.473 | |
| Within-R-squared | | | | | 0.343 |
| Between-R-squared | | | | | 0.362 |
| Overall-R-squared | | | | | 0.403 |

Standard errors in brackets
\*\*\* p<0.01, \*\* p<0.05, \* p<0.1

Table 11 outlines the country-level CART approach, using a regression-tree to replicate regressions outlined in Table 10. Overall, goodness-of-fit is roughly comparable to Table 10 OLS and fixed-effects regression-results. It does however grant insight via a more nuanced mapping of the causal structure, demonstrating that business-cycle and cash-on-market are more valuation-relevant for lower-revenue startups, while sectoral-beta appears to be more-relevant for higher-revenue startups. Overall, output-gap is demonstrated to be the second most-influential explanatory-variable, behind firm-revenue, and ahead of all risk-metrics.

*Table 11: Country-level CART*

| | | | | |
|---|---|---|---|---|
| OBS: | 1045 | | | |
| End Nodes: | 11 | | | |
| Complexity parameter | No. of Split | RMSE | Crossvalidation error | Crossvalidation St. Dev. |
| 0.31393071 | 0 | 1.0000000 | 1.0016062 | 0.03169662 |
| 0.14568789 | 1 | 0.6860693 | 0.6838544 | 0.03029474 |
| 0.02155412 | 2 | 0.5403814 | 0.5561247 | 0.02830774 |
| 0.01876351 | 3 | 0.5188273 | 0.5304358 | 0.02768721 |
| 0.01836602 | 4 | 0.5000638 | 0.5242283 | 0.02761847 |
| 0.01534326 | 5 | 0.4816977 | 0.5112754 | 0.02742454 |
| 0.01465354 | 6 | 0.4663545 | 0.5061729 | 0.02744926 |
| 0.01123159 | 7 | 0.4517009 | 0.4841989 | 0.02715784 |
| 0.01048947 | 8 | 0.4404694 | 0.4790536 | 0.02701238 |
| 0.01038662 | 9 | 0.4299799 | 0.4766123 | 0.02698313 |
| 0.01000000 | 10 | 0.4195933 | 0.4755234 | 0.02701681 |
| Variable Importance | | | | |
| Ln_Revenue | Output Gap | Ln_Beta | Country-Risk Premium | Cash-on-Market |
| 40 | 29 | 14 | 10 | 7 |

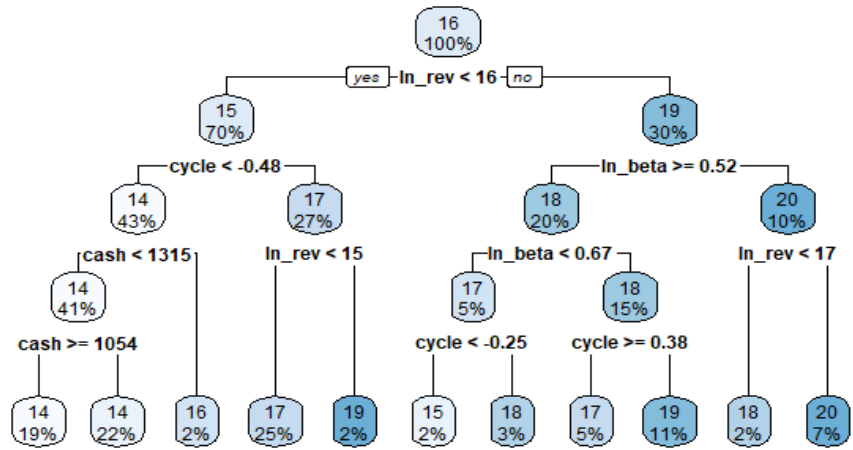

Table 12 examines the same relationship using country-level microtargeting, relying on country-categoricals rather than distinct country-level variables. Overall, Table 12 demonstrates explanatory-power to be comparable to that of Tables 10 and 11 OLS and CART regression using country-level economic-indicator, while outperforming those of Table 10's fixed-effects approach. Additionally, the dataset's country-categorical ranks second in terms of variable-importance, behind revenue. This outranks all risk-adjusted discount-factor determinants. Stated otherwise, country-level microtargeting demonstrates that country-categoricals have stronger explanatory-power than discount-factor determinants.

*Table 12: Country-level CART Microtargeting*

| | | | | |
|---|---|---|---|---|
| OBS: | 1045 | | | |
| End Nodes: | 15 | | | |
| Complexity parameter | No. of Split | RMSE | Crossvalidation error | Crossvalidation St. Dev. |
| 0.30592475 | 0 | 1.0000 | 1.0015530 | 0.03163665 |
| 0.08474915 | 1 | 0.6941 | 0.7090591 | 0.03011185 |
| 0.02283395 | 2 | 0.6093 | 0.6369268 | 0.02805479 |
| 0.02024432 | 3 | 0.5865 | 0.6364982 | 0.02885804 |
| 0.01604568 | 4 | 0.5662 | 0.6305592 | 0.02876314 |
| 0.01513984 | 5 | 0.5502 | 0.6355422 | 0.02893389 |
| 0.01410708 | 6 | 0.5351 | 0.6362563 | 0.02920216 |
| 0.01364862 | 7 | 0.5210 | 0.6260981 | 0.02885942 |
| 0.01356276 | 9 | 0.4937 | 0.6236566 | 0.02883585 |
| 0.01004333 | 10 | 0.4801 | 0.6017074 | 0.02796945 |
| 0.01000000 | 14 | 0.4399 | 0.5814898 | 0.02699292 |
| Variable Importance | | | | |
| Ln_Revenue | Country | Ln_Beta | Country-Risk Premium | |
| 45 | 27 | 14 | 13 | |

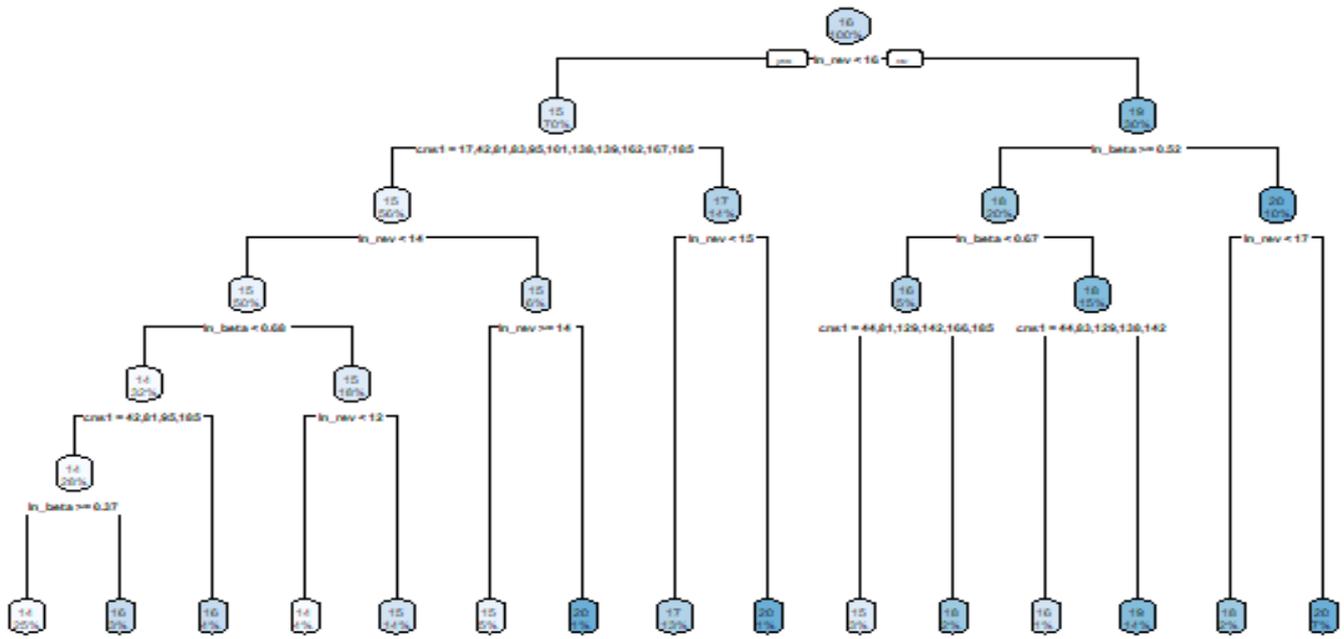

## The Municipal Level

While classical financial and economic views describe valuations as driven by concrete, measurable economic factors, (Fama, 1970) tied to revenue, risk, and industry-level indicators, and country-level macroeconomic indicators (Damodaran, 2009), financial media and industry-sources often tie valuations to city-level indicators. Within the literature, both Porter (1990), and Nathan and Vandore (2014) describe the economic importance of specific cities and of local-level impacts, attributing the economic-impact to complex competitive, cooperative, and interactive relationships among suppliers, consumers, competitors, and networks.

Table 13 uses OLS and fixed-effects to describe valuation-impacts of city-level indicators, counterintuitively finding that many city-level indicators, ranging from deal-density to physical-infrastructure, to city-level intellectual-property to have negative valuation-impacts for startups. While counterintuitive at first-glance, this may indicate that deal-dense, IP-dense, high-GDP cities represent a challenging competitive-environment for startups, in an environment that might be characterized by potentially-limited availability of business angel and venture-capital funding.

*Table 13: City-level OLS and Fixed Effects*

**City-level Regressions**

| VARIABLES | (1) Ln_Valuation | (2) Ln_Valuation | (3) Ln_Valuation | (4) Ln_Valuation | (5) Ln_Valuation | (6) Ln_Valuation | (7) Ln_Valuation | (8) Ln_Valuation |
|---|---|---|---|---|---|---|---|---|
| Ln_Revenue | 0.6514*** | 0.6044*** | 0.6374*** | 0.6408*** | 0.6286*** | 0.6432*** | 0.5970*** | 0.6022*** |
|  | [0.034] | [0.046] | [0.036] | [0.037] | [0.033] | [0.037] | [0.044] | [0.037] |
| Ln_Beta | -1.0886*** | -1.0775** | -1.0739*** | -0.9750** | -1.3881*** | -0.9119** | -1.4431*** | -1.3550*** |
|  | [0.388] | [0.473] | [0.411] | [0.413] | [0.378] | [0.420] | [0.451] | [0.401] |
| Country Risk Premium | -38.3685*** | -11.0416 | -52.9071*** | -52.4256*** | -35.8853*** | -51.6485*** | 0.8163 | -43.0762* |
|  | [13.092] | [26.653] | [16.565] | [16.842] | [12.694] | [16.720] | [25.667] | [23.707] |
| Municipal GDP |  | -0.1247*** |  |  |  |  | -0.1669** |  |
|  |  | [0.031] |  |  |  |  | [0.071] |  |
| City Physical Infrastructure |  |  | -0.0157*** |  |  |  | - |  |
|  |  |  | [0.005] |  |  |  | - |  |
| City Intellectual Property |  |  |  | -0.0000 |  |  | 0.0000*** |  |

|  |  |  |  |  | [0.000] |  | [0.000] |  |
|---|---|---|---|---|---|---|---|---|
| No. of deals in the city |  |  |  |  |  | -0.0105*** |  | -0.0123*** |
|  |  |  |  |  |  | [0.002] |  | [0.002] |
| City Productivity |  |  |  |  |  |  | 0.0000 | 0.0000 |
|  |  |  |  |  |  |  | [0.000] | [0.000] |
| Constant | 7.9171*** | 13.5138*** | 8.8253*** | 8.2471*** | 8.8485*** | 7.8635*** | 14.6359*** | 9.0680*** |
|  | [0.615] | [1.647] | [0.685] | [0.707] | [0.613] | [0.825] | [2.043] | [0.695] |
| Observations | 646 | 393 | 541 | 541 | 646 | 541 | 393 | 588 |
| R-squared | 0.41 | 0.47 | 0.42 | 0.42 | 0.45 | 0.42 | 0.53 |  |
| Adjusted R-squared | 0.408 | 0.461 | 0.421 | 0.411 | 0.444 | 0.412 | 0.524 |  |
| Within-R-squared |  |  |  |  |  |  |  | 0.325 |
| Between-R-squared |  |  |  |  |  |  |  | 0.410 |
| Overall-R-squared |  |  |  |  |  |  |  | 0.418 |

Standard errors in brackets
a p<0.01, b p<0.05, c p<0.1

### City-level Indicators CART

Table 11 replicates the city-level OLS and fixed-effects regressions described in Table XX. Compared to OLS and fixed-effects regressions, the city-level CART regression tree's lowest branches have goodness-of-fit indicators roughly-equivalent to the city-level fixed-effects regression, but are outperformed in terms of explanatory-power by the city-level OLS joint-indicator regression. Nevertheless, the CART approach does grant us novel insight, given the fault-lines and variable-rankings demonstrated. In terms of explanatory-power, city-level deal-density dramatically outweighs both risk-premium factors and other city-level indicators. Additionally, we see that both extremely low-valuation and high-valuation deals are concentrated deal-dense cities. This relationship explains and contextualizes both negative regression-coefficient expressed by Number of Deals in City in Table 10, as well as the larger goodness-of-fit compared to other regressions in Table 10.

*Table 14: City-level Indicators CART*

| OBS: | 1045 |  |  |  |
|---|---|---|---|---|
| End Nodes: | 13 |  |  |  |
| Complexity parameter | No. of Split | RMSE | Crossvalidation error | Crossvalidation St. Dev. |
| 0.31393064 | 0 | 1.0000 | 1.0008855 | 0.03164963 |
| 0.05684246 | 1 | 0.6861 | 0.6998654 | 0.02970526 |
| 0.02469787 | 2 | 0.6292 | 0.6526063 | 0.02862365 |
| 0.02019285 | 5 | 0.5551 | 0.6263861 | 0.02933202 |
| 0.01513657 | 8 | 0.4946 | 0.5480737 | 0.02702584 |
| 0.01496123 | 9 | 0.4794 | 0.5350114 | 0.02643029 |
| 0.01220708 | 10 | 0.4645 | 0.5243352 | 0.02578351 |
| 0.01206923 | 11 | 0.4522 | 0.5160791 | 0.0252533 |
| 0.01000000 | 12 | 0.4402 | 0.5012024 | 0.02509131 |

Variable Importance

| Ln_Revenue | No. of deals in the city | Ln_Beta | Country-Risk Premium | City Intellectual Property | City Productivity | City Physical Infrastructure |
|---|---|---|---|---|---|---|
| 49 | 21 | 11 | 7 | 6 | 5 | 1 |

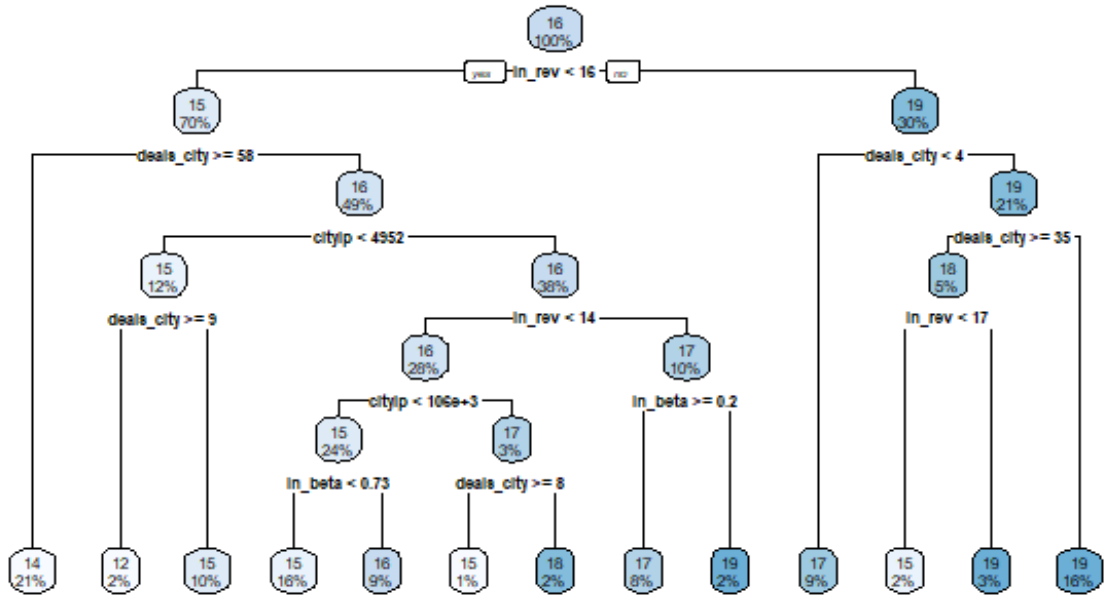

## City-level CART Microtargeting

In contrast to the city-level indicators CART approach outlines in Table 14, the city-level microtargeting approach tells a somewhat more detailed story, despite having similar goodness-of-fit indicators and fewer indicators to work with in order to capture the relationship. Table 12 outlines the CART regression tree for city-level microtargeting. In comparison to the city CART approach, the city categorical variable is nearly as powerful as firm-revenue, while the regression-tree as a whole has stronger goodness-of-fit at its lowest branches than either the city-indicators CART tree or the city-level regressions table. According to this tree's indications, which city a given startup is located in predicts valuation more accurately than industry-level and country-level risk-factors do. Additionally, high-valuation startups appear to be concentrated in a relatively-small number of cities, as indicated by the lower-branches in high-valuation segment of the valuation-distribution.

*Table 15: City-level CART Microtargeting*

| Complexity parameter | No. of Split | RMSE | Crossvalidation error | Crossvalidation St. Dev. |
|---|---|---|---|---|
| OBS: 1045 | | | | |
| End Nodes: 14 | | | | |
| 0.29615290 | 0 | 1.0000 | 1.0056621 | 0.03186467 |
| 0.08350413 | 1 | 0.7038 | 0.7132971 | 0.02923645 |
| 0.03857233 | 2 | 0.6203 | 0.6563125 | 0.02665414 |
| 0.01917984 | 3 | 0.5818 | 0.6361755 | 0.02795105 |
| 0.01828399 | 4 | 0.5626 | 0.6362341 | 0.02828942 |
| 0.01438177 | 6 | 0.5260 | 0.6177926 | 0.02766236 |
| 0.01268354 | 9 | 0.4829 | 0.5759536 | 0.02631319 |
| 0.01124848 | 10 | 0.4702 | 0.5574863 | 0.02679365 |
| 0.01082355 | 11 | 0.4589 | 0.5570876 | 0.02680696 |
| 0.01000000 | 13 | 0.4373 | 0.5515336 | 0.02678294 |

| Variable Importance | | | |
|---|---|---|---|
| Ln_Revenue | City | Ln_Beta | Country-Risk Premium |
| 43 | 31 | 16 | 11 |

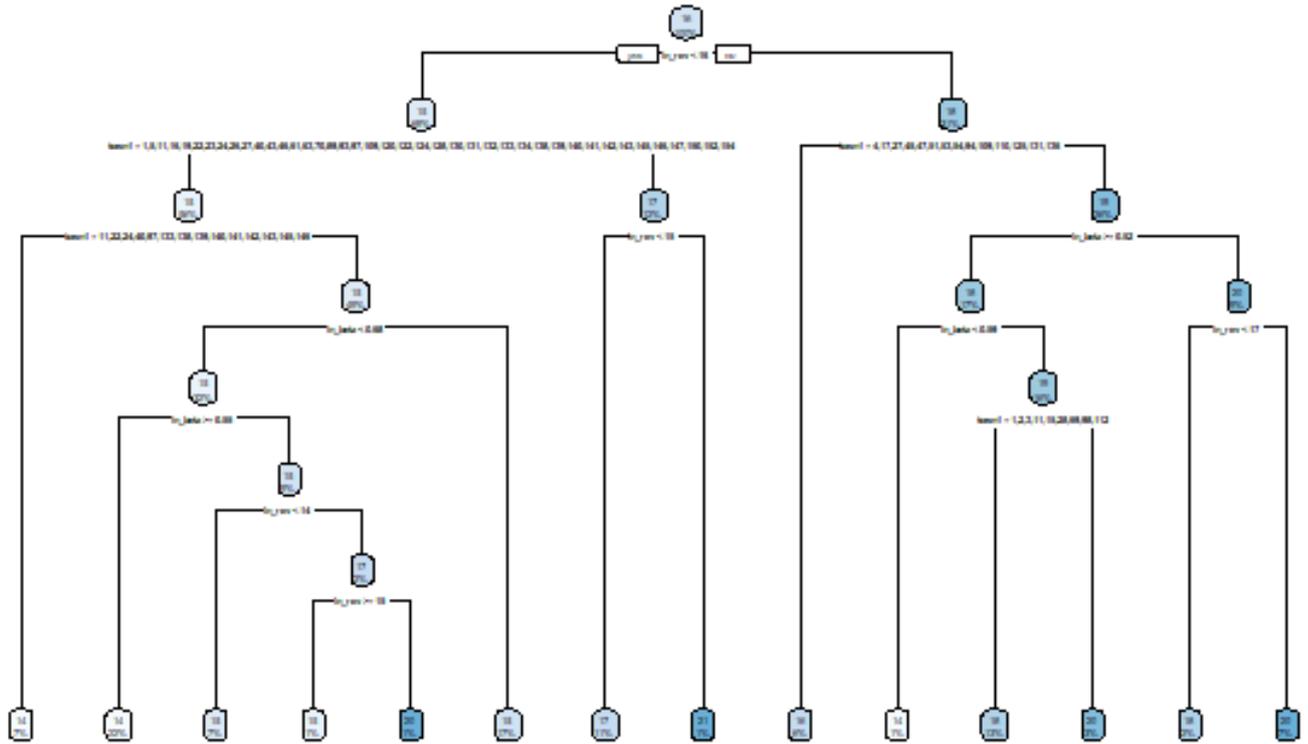

**Firm-Ownership**

In this study's dataset, a firm can be privately-held, held by shareholders, or be owned as a subsidiary. While economic theory holds that ownership-structure can be complex (Berre and Le Pendeven, 2022) and also impact startup-valuation (Gornall and Strebulaev, 2020). Privately-held firms can have investment-clauses aimed at investor-protection, ranging from founder-vesting to exit, leaver, and tag-along clauses aimed at maximizing both investor-protection and entrepreneur incentive-alignment. Meanwhile, subsidiaries might have access to parent-company-affiliated support (Chemmanur et al., 2014).

Nevertheless, OLS and fixed-effects regressions outlined in Table 16, which examines dummy variables for private-ownership and subsidiary-ownership demonstrate that only the for private-ownership significantly impacts startup-valuations, granting a valuation-premium, while subsidiaries and publicly-traded firms see no significant valuation-impact. This is followed by a fixed-effects regression, modelling ownership-type as a categorical variable rather than its constituent dummy-variables. Overall, the regression including the privately-held firm dummy demonstrates both an improvement in goodness-of-fit compared to the baseline DCF-factor regression, as well as a statistically-significant dummy-coefficient, whereas the publicly-traded and subsidiary dummies have no impact on the regression, while the firm-ownership fixed-effects regression demonstrates substantial goodness-of-fit underperformance compared to the DCF-factor regression, as well as all other OLS regressions. The ownership-type fixed-effects regression in contrast, demonstrates weaker goodness-of-fit than not only the dummy-variable-controlling OLS regressions, but DCF-factor regression as well.

*Table 16: Impact of Ownership-type OLS and Fixed-Effects*

**Valuation and Startup-Ownership**

| VARIABLES | (1) Ln_Valuation | (2) Ln_Valuation | (3) Ln_Valuation | (4) Ln_Valuation | (5) Ln_Valuation |
|---|---|---|---|---|---|
| Ln_Revenue | 0.6514*** | 0.5596*** | 0.6574*** | 0.6510*** | 0.3586*** |
|  | [0.034] | [0.034] | [0.035] | [0.034] | [0.062] |
| Ln_Beta | -1.0887*** | -1.4991*** | -1.0544*** | -1.1015*** | -3.0419*** |
|  | [0.388] | [0.368] | [0.389] | [0.388] | [0.598] |
| Country Risk Premium | -38.3653*** | -38.1861*** | -38.4168*** | -37.5781*** | -10.2912 |
|  | [13.092] | [12.347] | [13.092] | [13.146] | [20.784] |
| Privately-Held |  | 1.7839*** |  |  |  |
|  |  | [0.198] |  |  |  |
| Publicly-traded |  |  | -0.6444 |  |  |
|  |  |  | [0.648] |  |  |
| Subsidiary |  |  |  | 0.6034 |  |
|  |  |  |  | [0.869] |  |
| Constant | 7.9170*** | 9.0141*** | 7.8222*** | 7.9169*** | 14.6366*** |
|  | [0.615] | [0.593] | [0.623] | [0.616] | [1.129] |
| Observations | 646 | 646 | 646 | 646 | 199 |
| R-squared | 0.41 | 0.48 | 0.41 | 0.41 |  |
| Adjusted R-squared | 0.408 | 0.473 | 0.408 | 0.407 |  |
| Number of statuses |  |  |  |  | 3 |
| Within-R-squared |  |  |  |  | 0.290 |
| Between-R-squared |  |  |  |  | 0.055 |
| Overall-R-squared |  |  |  |  | 0.282 |

Standard errors in brackets
*** p<0.01, ** p<0.05, * p<0.1

Although this study's dataset does not have smaller indicators tied to ownership structure, the dataset does contain categoricals for ownership itself, making an ownership-driven microtargeting of startup-valuation possible. In contrast to all other categorical variables included in this study, ownership does not make an appearance on the CART regression-tree results outlined in Table 17. Nevertheless, while OLS findings outlined in Table 16 are mostly corroborated by Table 17's microtargeting-approach, the CART regression-tree demonstrates substantially-stronger goodness-of-fit than either the OLS or fixed-effects approaches. Because the firm-ownership categorical is excluded from CART results, this might, in principle, be due to elaboration added by mapping the regression-tree's variable-hierarchy.

*Table 17: Impact of Ownership-type CART Microtargeting*

| OBS: | 1045 | | | |
|---|---|---|---|---|
| End Nodes: | 10 | | | |
| Complexity parameter | No. of Split | RMSE | Crossvalidation error | Crossvalidation St. Dev. |
| 0.31393064 | 0 | 1.0000 | 1.0035362 | 0.03175907 |
| 0.03811921 | 1 | 0.6861 | 0.694355 | 0.02966098 |
| 0.0233742 | 2 | 0.6480 | 0.6667157 | 0.0293431 |
| 0.02155411 | 3 | 0.6246 | 0.6466988 | 0.02798392 |
| 0.01876349 | 4 | 0.6030 | 0.6369146 | 0.02771042 |
| 0.01534327 | 5 | 0.5843 | 0.6212674 | 0.02727309 |
| 0.01450141 | 6 | 0.5689 | 0.6185398 | 0.02710567 |
| 0.01152477 | 8 | 0.5399 | 0.5867917 | 0.02587693 |
| 0.01152093 | 9 | 0.5284 | 0.5763595 | 0.02552331 |
| 0.01000000 | 11 | 0.5053 | 0.5739194 | 0.02631373 |

| Variable Importance | | |
|---|---|---|
| Ln_Revenue | Country-Risk Premium | Ln_Beta |
| 59 | 23 | 18 |

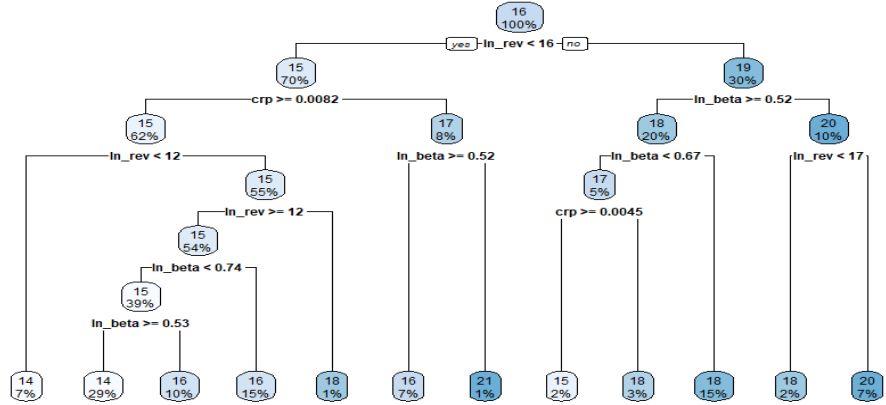

## Business Model

Beyond revenues, risk-factors, and geographic-factors, business-models, outlined in terms of customer-focus also have the potential to play an important deterministic role in determining startup-valuations. Among practitioner-models, both the Payne Scorecard Model and the Berkus Model make explicit mention of details of the business model (Payne, 2011; Berkus, 2016). In principle, business-models can impact firm-level revenue-models, growth-rates, business-networks, and risk-profiles, all of which can have a direct impact on the startup's firm-valuation (Damodaran, 2009, Berre and Le Pendeven 2022). Table 18 demonstrates the OLS and fixed-effects regressions capturing business models. In principle, the business-model categorical-variable describes whether the firm's business model is business-to-business (B2B), business-to-consumer (B2C), business-to-government (B2G), business-to-business and consumer (B2B&C). Here, two approaches are taken to incorporate business models focus. The OLS regressions include dummies for B2B, B2C, and B2B&C, while the fixed-effects regression includes business-model fixed-effects.

*Table 18: Business Models and Valuation: OLS and Fixed-effects Regressions*

**Business Models and Valuation: OLS and Fixed-effects Regressions**

| VARIABLES | (1) Ln_Valuation | (2) Ln_Valuation | (3) Ln_Valuation | (4) Ln_Valuation | (5) Ln_Valuation |
|---|---|---|---|---|---|
| Ln_Revenue | 0.6514*** | 0.6498*** | 0.6006*** | 0.6381*** | 0.5228*** |
|  | [0.034] | [0.034] | [0.032] | [0.035] | [0.042] |
| Ln_Beta | -1.0887*** | -1.1630*** | -1.6762*** | -1.0276*** | -2.2538*** |
|  | [0.388] | [0.393] | [0.364] | [0.389] | [0.473] |
| Country Risk Premium | -38.3653*** | -37.4777*** | -39.4704*** | -37.0535*** | -3.3346 |
|  | [13.092] | [13.111] | [12.131] | [13.099] | [14.886] |
| B2B |  | 0.2873 |  |  |  |
|  |  | [0.248] |  |  |  |
| B2C |  |  | 2.4084*** |  |  |
|  |  |  | [0.233] |  |  |
| B2B&C |  |  |  | 0.4037* |  |
|  |  |  |  | [0.245] |  |
| Constant | 7.9170*** | 7.9310*** | 8.6406*** | 7.9960*** | 11.1678*** |
|  | [0.615] | [0.615] | [0.575] | [0.617] | [0.772] |
| Observations | 646 | 646 | 646 | 646 | 318 |
| R-squared | 0.41 | 0.41 | 0.49 | 0.41 |  |
| Adjusted R-squared | 0.408 | 0.408 | 0.491 | 0.409 |  |
| Number of Business-Models |  |  |  |  | 3 |
| Within-R-squared |  |  |  |  | 0.448 |
| Between-R-squared |  |  |  |  | 0.160 |
| Overall-R-squared |  |  |  |  | 0.412 |

Standard errors in brackets
*** p<0.01, ** p<0.05, * p<0.1

As is the case with Tables 16 and 17 examining firm-ownership, this study's dataset does not have smaller indicators tied to business-models, although the dataset does contain categoricals for business-model itself, making business-model-driven microtargeting of startup-valuation possible. As is the case with the firm-ownership tables, Tables 18 and 19 compare business-model fixed-effects regression to OLS including business-model dummies, to CART-based microtargeting using a business-model categorical variable. Compared to Table 18, the microtargeting approach outlined in Table 19 demonstrates that while business-model customer-focus boasts the model's second-highest variable-importance, ahead of the model's discount-factor determinants, and behind revenue, business-model customer-focus also plays its primary deterministic-role in low-revenue, low-valuation firms, where business-model customer-focus is interactive with firm-revenue, suggesting the valuation might be subject to a somewhat-complex functional-form, which indicates that startups using B2B business-models may experience a valuation-premium. This approach also finds stronger goodness-of-fit than either the OLS of fixed-effects models, an improvement in explanatory power, which can largely be attributed to the added-value drawn from mapping the model's variable-hierarchy.

*Table 19: Business Models and Valuation: CART Microtargeting*

| OBS: | 1045 | | | |
|---|---|---|---|---|
| End Nodes: | 10 | | | |
| Complexity parameter | No. of Split | RMSE | Crossvalidation error | Crossvalidation St. Dev. |
| 0.29613500 | 0 | 1.0000 | 1.0015277 | 0.03168565 |
| 0.10520043 | 1 | 0.7039 | 0.7009401 | 0.02998025 |
| 0.03004077 | 2 | 0.5987 | 0.6022224 | 0.02768759 |
| 0.02680278 | 3 | 0.5686 | 0.5974261 | 0.02762711 |
| 0.02048318 | 4 | 0.5418 | 0.5759336 | 0.02679744 |
| 0.01900543 | 5 | 0.5213 | 0.5750794 | 0.02685839 |
| 0.01340014 | 6 | 0.5023 | 0.5561843 | 0.02579164 |
| 0.01131482 | 7 | 0.4889 | 0.5515973 | 0.02620067 |
| 0.01028209 | 8 | 0.4776 | 0.5379973 | 0.02603295 |
| 0.01000000 | 9 | 0.4673 | 0.5249817 | 0.02572137 |

| Variable Importance | | | |
|---|---|---|---|
| Ln_Revenue | Customers | Country-Risk Premium | Ln_Beta |
| 47 | 25 | 14 | 14 |

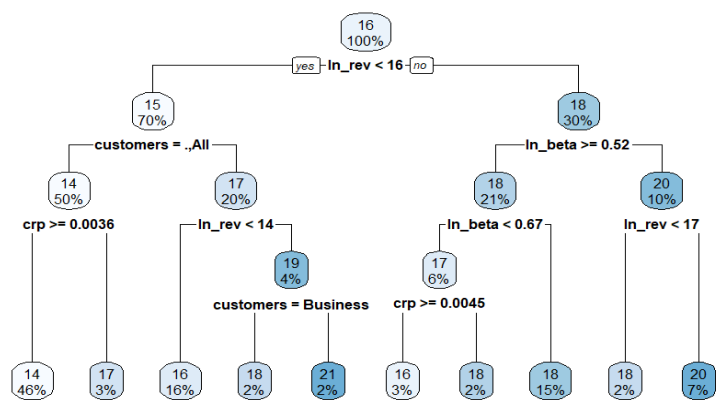

### Joint-Fixed-Effects and Combined Microtargeting

In order to compare and combine explanatory-power of all available categorical-variables, as individual and joint fixed-effects, Table 20 compares the discounted-cashflow approximation-model outlined in Table 4 to structurally-similar fixed-effects models capturing individual and joint-fixed-effects. This includes fixed-effects using categorical-variables for industry, country, city, investor-type, business-model, and ownership-status.

*Table 20: DCF-based Regressions in OLS, Fixed-Effects, and Joint Fixed-Effects*

**DCF-based Regressions in OLS, Fixed-Effects, and Joint Fixed Effects**

| VARIABLES | OLS Ln_Valuation | Sector FE Ln_Valuation | Country FE Ln_Valuation | City FE Ln_Valuation | Invest. Type FE Ln_Valuation | Business-Model FE Ln_Valuation | Owner. Status FE Ln_Valuation | Joint FE Ln_Valuation |
|---|---|---|---|---|---|---|---|---|
| Ln_Revenue | 0.6514*** | 0.6318*** | 0.5817*** | 0.6022*** | 0.4049*** | 0.5228*** | .321037*** | 0.2551*** |
|  | [0.034] | [0.034] | [0.034] | [0.037] | [0.034] | [0.042] | [0.073] | [0.096] |
| Ln_Beta | -1.0886*** | -0.3805 | -1.5274*** | -1.3550*** | -1.1420*** | -2.2538*** | -4.086*** | -2.5047** |
|  | [0.388] | [0.578] | [0.379] | [0.401] | [0.363] | [0.473] | [0.724] | [1.244] |
| Country Risk Premium | -38.3685*** | -38.3666*** | -6.3079 | -43.0762* | -24.6074* | -3.3346 | -17.779 | -42.969 |
|  | [13.092] | [12.737] | [21.354] | [23.707] | [12.611] | [14.886] | [21.589] | [35.294] |
| Constant | 7.9171*** | 7.6960*** | 9.3378*** | 9.0680*** | 11.4272*** | 11.1678*** | 16.002*** | 15.765*** |
|  | [0.615] | [0.690] | [0.691] | [0.695] | [0.693] | [0.772] | [1.299] | [15.765] |
| Observations | 646 | 643 | 646 | 588 | 528 | 318 | 199 | 140 |
| R-squared | 0.41 |  |  |  |  |  |  |  |
| Adjusted R-squared | 0.408 |  |  |  |  |  |  |  |
| Number of categories |  | 18 | 20 | 58 | 17 | 3 | 3 | 90 |
| Within-R-squared |  | 0.356 | 0.343 | 0.325 | 0.253 | 0.448 | 0.290 | 0.073 |
| Between-R-squared |  | 0.691 | 0.362 | 0.410 | 0.619 | 0.160 | 0.0550 | 0.294 |
| Overall-R-squared |  | 0.409 | 0.403 | 0.418 | 0.399 | 0.412 | 0.282 | 0.282 |

Standard errors in brackets
*** p<0.01, ** p<0.05, * p<0.1

Overall, Table 20's regressions, all of which have DCF-consistent directional-signs, feature statistically-significant coefficients for revenue and sectoral-beta, while country-risk premium's negative coefficient is significant in four of eight regressions, indicating in principle that country-risk premium's valuation-impact is likely overshadowed by the explanatory-power of some of the categorical factors used to construct Table 20's fixed-effects models, such as the country-categorical or joint-fixed-effects.

While city and business-model fixed-effects boast the strongest goodness-of-fit indicators, the fixed-effects models in Table 20 generally demonstrate loss of explanatory-power as the number of fixed-effect categories increases, with the joint-fixed effects regression, boasting 90 fixed-effects categories, having the lowest explanatory-power of any model, while also risking running into sample-size issues, which may wholly undermine its validity due to per-category number-of-observations, as described by Wooldridge (2010) and Theall et al., (2011).

On the other hand, Table 21 approaches joint-categoricals via CART, while presenting strong goodness-of-fit for its valuation-model, as well as further insight on key fault lines and relative hierarchical-ranking of valuation-factors. In addition to not suffering from loss of explanatory-power due to the increases of the number of categories as is the case with joint fixed-effects, use of categorical-variable-based microtargeting has the additional advantage of generally having fewer missing values. Compared to regression-trees which use only one categorical-variable, which mostly place the categorical valuation-determinant ahead of discount-factors sectoral-beta and country-risk-premium, but behind firm-revenues in terms of variable-importance, Table 21's regression-tree places all of the model's categorical valuation-determinants ahead of both revenue and the discount-factors, effectively describing that the categorical valuation-determinants are more information-dense (or at least have more substantial explanatory-power) than all of model's numerical valuation-determinants. Overall, the regression-tree describes that while investor-type is the most influential valuation-determinant, higher-valuation startups are evaluated on the basis of revenue, and industry-risk, while lower-valuation startups see their valuations depend mainly city-location. Table 21's regression-tree dendrogram demonstrates that while higher-valuation startups have valuations driven by revenue, as well as by customer-focus business-model and sectoral-beta, lower-valuation startups have values driven primarily by geographical-factors.

*Table 21: CART Microtargeting: Valuation using Industry, Country, City, Investor-type, Business-model, and Ownership-status*

| OBS: | 1045 | | | |
|---|---|---|---|---|
| End Nodes: | 10 | | | |
| Complexity parameter | No. of Split | RMSE | Crossvalidation error | Crossvalidation St. Dev. |
| 0.3614 | 0 | 1.0000 | 1.0017 | 0.0317 |
| 0.0638 | 1 | 0.6386 | 0.6530 | 0.0287 |
| 0.0518 | 2 | 0.5747 | 0.6246 | 0.0283 |
| 0.0492 | 3 | 0.5230 | 0.5810 | 0.0275 |
| 0.0206 | 4 | 0.4738 | 0.5154 | 0.0236 |
| 0.0178 | 6 | 0.4326 | 0.4978 | 0.0234 |
| 0.0133 | 7 | 0.4148 | 0.4868 | 0.0231 |
| 0.0113 | 8 | 0.4015 | 0.4792 | 0.0233 |
| 0.0100 | 9 | 0.3903 | 0.4795 | 0.0236 |

Variable Importance

| Investor Type | Country | City | Customer | Sector | Ln_Revenue | Country-Risk Premium | Ln_Beta |
|---|---|---|---|---|---|---|---|
| 28 | 17 | 16 | 13 | 11 | 9 | 4 | 2 |

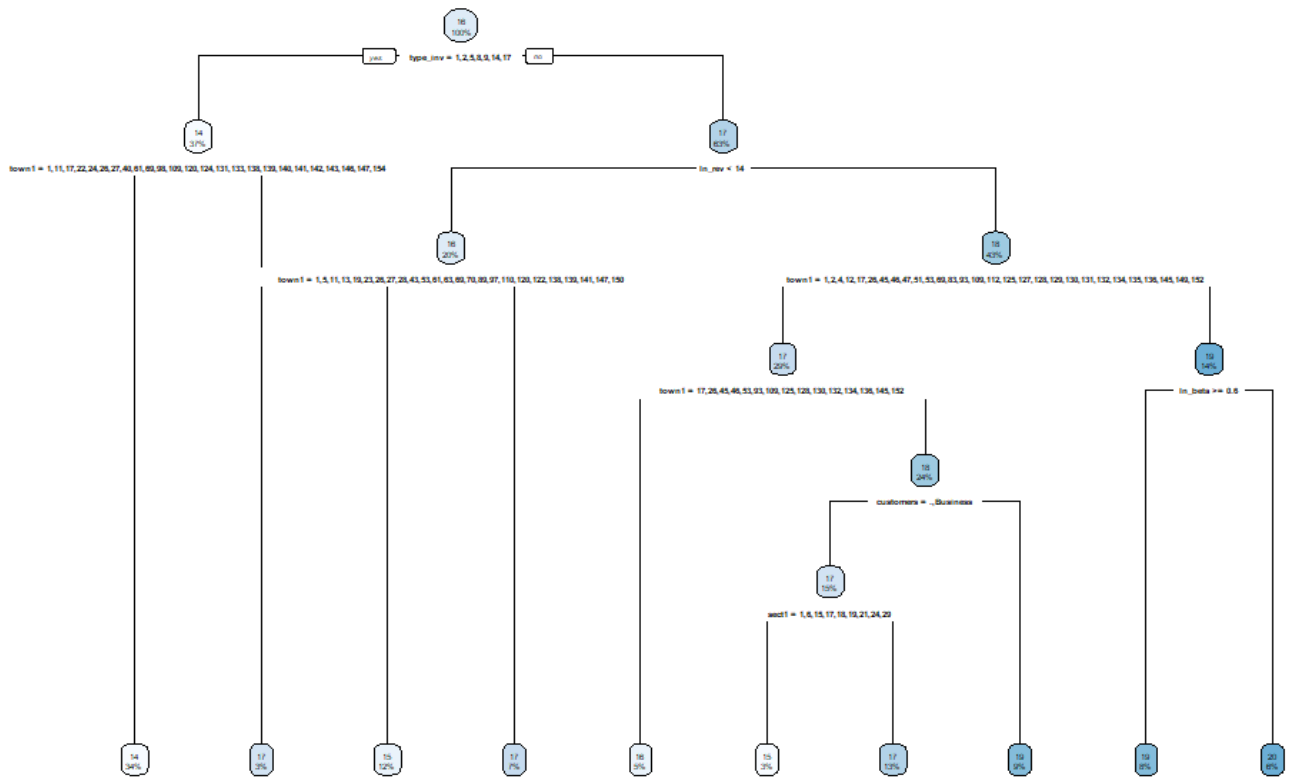

Because Table 21 describes the joint-categorical microtargeting-model as driven primarily by investor-type, country, city, customer-type, and sector, the microtargeting-tree can be re-executed omitting the DCF-factors – the model's only numerical factors. – entirely. Table 22 demonstrates the microtargeting-tree approach, estimating valuation on the basis of microtargeting alone. Compared to both the fixed-effects regression-models in Table 20, and Table 21's the joint-categorical CART model simulating the last regression in Table 20, goodness-of-fit is roughly comparable to both Table 22's goodness-of-fit, as well as the more powerful fixed-effects regressions in Table 20. These results not only confirm H3, but also managed to do so without reliance on sometimes difficult-to-obtain numerical DCF-factors, relying only on categorical variables instead.

*Table 22: CART Microtargeting: Valuation using only categoricals: Industry, Country, City, Investor-type, Business-model, and Ownership-status*

| OBS: | 1045 | | | |
|---|---|---|---|---|
| End Nodes: | 10 | | | |
| Complexity parameter | No. of Split | RMSE | Crossvalidation error | Crossvalidation St. Dev. |
| 0.3614 | 0 | 1.0000 | 1.0019 | 0.0317 |
| 0.0796 | 1 | 0.6386 | 0.6531 | 0.0286 |
| 0.0557 | 2 | 0.5589 | 0.5883 | 0.0281 |
| 0.0326 | 3 | 0.5032 | 0.5363 | 0.0243 |
| 0.0163 | 5 | 0.4380 | 0.4888 | 0.0232 |
| 0.0155 | 6 | 0.4217 | 0.4856 | 0.0237 |
| 0.0134 | 7 | 0.4062 | 0.4704 | 0.0243 |
| 0.0133 | 8 | 0.3929 | 0.4645 | 0.0243 |
| 0.0100 | 9 | 0.3796 | 0.4473 | 0.0233 |
| Variable Importance | | | | |
| Investor Type | Country | City | Customer | Sector |
| 28 | 17 | 16 | 13 | 11 |

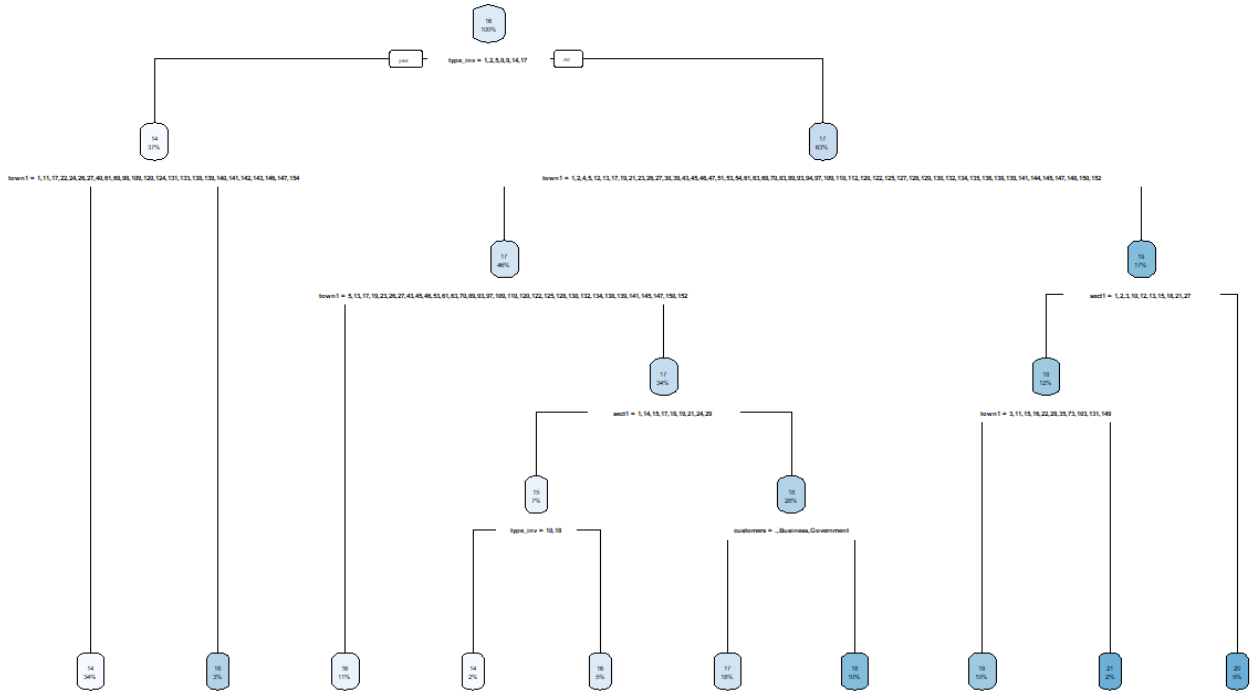

Compared to Table 21, goodness-of-fit indicators in Table 22 actually demonstrate stronger explanatory-power, despite having fewer explanatory-factors. As is the case in Table 21, Table 22 demonstrates that valuation is driven by investor-type, followed by country, city, customer-type, and sector. Overall, the regression-tree describes that while investor-type is the most influential valuation-determinant, lower-valuation startups see their valuations depend mainly city-location, while higher-valuation startups have valuations driven by city-location and industry.

## 6. Discussion and Conclusion

The results presented in this study demonstrate very powerful findings. Essentially, this study demonstrates that it is possible to estimate startup-valuation via microtargeting, and on the basis of accumulated categorical-data. For this study's microtargeting approach, categorical valuation-factors are compared in terms of explanatory-power primarily to numerical valuation-factors which they could viably act as substitutes for, such as the country-categorical-factor in place of country-level business cycle and cash-on-market, and city-categorical-factor in place of municipal GDP and city-level IP.

This study finds that not only does the CART-based regression-tree approach either match or outperform OLS in terms of goodness-of-fit, but also that this approach allows use of categorical-based microtargeting, which enables us to make accurate valuation-predictions even when faced with the non-availability of firm-specific performance-metrics which classical firm-valuation models rely on, such as firm-revenue, net-income, CAPEX, or balance-sheet data. This fact makes such an approach extremely valuable in startup-markets, where much of the relevant firm-level data is kept confidential or protected by non-disclosure agreements.

The general CART-model approach value-added can largely be attributed to the value-added by mapping the fault-lines and variable-importance hierarchy, which adds a layer of complexity to empirical findings which can not only be of immediate accountable use in markets, and is able to contextualized precisely why and how scorecard models such as those of Payne (2011) and Berkus (2016) enjoy widespread use among practitioners. Additionally, the variable-selection flexibility that the CART-approach allows grants opportunity to model, incorporate, and account-for a wide-range of contextual-factors – factors which both peer-review literature and practitioner-literature consider deterministic and pivotal (Berre and Le Pendeven, 2022) – into valuation models. Essentially, by applying scorecard-based valuation-approaches and segmented valuation-model approaches, economists, investors and market-practitioners are de-facto applying hierarchical segmented valuation-approaches because hierarchical algorithm approaches such as CART, AHC, and random forest result in segmented scorecard-style valuation-models.

This paper's findings provide critical implications for several interested parties. First, for market-practitioners, this study's findings not only validate and contextualize the use of segmented and scorecard-based valuation-approaches, but also provide blueprint for the further incorporation of contextual and qualitative valuation-factors. Second, for startups and entrepreneurs, a solid-grasp on CART-model insights on hierarchical-functional-forms and key-fault lines can help optimally-place startups in order to maximize valuations given sometimes-complex investor-negotiations. Meanwhile, outsiders and third-parties, faced with unavailability of firm-level performance and asset-data due to privacy, non-disclosure, and measurement-difficulty of intangible-assets, this paper's microtargeting approach provides a viable methodological bypass to this challenge, given that microtargeting provides a viable way to reach actionable valuation-predictions independently of firm-level performance data. Lastly, because this paper's CART-approach bypasses the loss of explanatory-power as categorical-factor-dimensionality increases, microtargeting would also make a suitable approach for AI-based automation, and the bid-data and IoT-industry grows and diversifies. Because large big-data-industry firms such as Google and Meta, who have played significant roles in the development and diffusion of AI and commercial-use of algorithms are also known to engage in corporate-entrepreneurship, the diffusion of valuation-algorithms in future startup-markets might someday be driven similar types of valuation-algorithms.

Future studies can be expected to expand this field by applying this methodological approach to other topics in entrepreneurial finance, such as startup-selection or modelling startup-survivorship. Furthermore, studies focusing on the theoretical modeling of hierarchical-modelling approaches for valuation, as well as for parallel-uses for hierarchical-model functional-forms within entrepreneurial-finance are also likely to arise in the near future.

# 7. References

Ang, Y. Q., Chia, A., and Saghafian, S. (2020) "Using Machine Learning to Demystify Startups Funding, Post-Money Valuation, and Success." HKS Faculty Research Working Paper Series RWP20-028, August 2020. Harvard Kennedy School.

Bai, S. and Zhao, Y. (2021). "Startup Investment Decision Support: Application of Venture Capital Scorecards Using Machine Learning Approaches". Systems. Vol. 9(3):55. July 2021

Berre, M. (2023), "Hierarchical and Segmented Approaches to Startup Valuation. What they are. Why they Work" Book Chapter.

Berre, M. and Le Pendeven, B. (2022), "What do we know about start-up valuation drivers? A systematic literature review". Venture Capital. DOI: 10.1080/13691066.2022.2086502.

Charnes, A., Cooper, W. W., and Rhodes, E., (1978). "Measuring the Efficiency of Decision Making Units". European Journal of Operational Research 2 (6), 429–444.

Chemmanur, T. J., Loutskina, E. & Tian, X. (2014). Corporate Venture Capital, Value Creation, and Innovation. *Review of Financial Studies*, Vol. 27, issue 8, 2434-2473

Damodaran, A. (2009). Valuing Young, Start-Up and Growth Companies: Estimation Issues and Valuation Challenges. SSRN Scholarly Paper ID 1418687. Rochester, NY: Social Science Research Network.

Damodaran, A. (2002). "Investment Valuation: Tools and Techniques for Determining the Value of Any Asset". 2nd ed. New York: Wiley, 2002.

Damodaran, A., Liu, C. H. (1993) "Insider Trading as a Signal of Private Information", The Review of Financial Studies, Vol. 6, No. 1, pp. 79-119, 1993

Eling, M. and Luhnen, M. (2010). "Frontier Efficiency Methodologies to Measure Performance in the Insurance Industry: Overview and New Empirical Evidence", The Geneva Papers on Risk and Insurance - Issues and Practice volume 35, pages 217–265(2010)

Ernst & Young (2020). "Startup Funding eGuide: A roadmap on how to raise capital as a startup". Startup Funding - EY - The Factory.

Huber, M., and Imhof, D. (2019) "Machine learning with screens for detecting bid-rigging cartels". International Journal of Industrial Organization Volume 65, July 2019, Pages 277-301

Fama, E. (1970). Efficient Capital Markets: A Review of Theory and Empirical Work. *The Journal of Finance*, Vol. 25, 383-417

Gompers, P. A. Gornall, W., Kaplan, S.A., and Strebulaev, I. A. (2020). How do venture capitalists make decisions? *Journal of Financial Economics*. Vol. 135 (2020) Pages 169–190

Gompers, P., and Lerner, J. (2001), The Money of Invention (Harvard Business School Press, Boston, MA).

Gompers, P. and Lerner, J. (1999), "The Venture Capital Cycle". MIT Press, Cambridge, MA.

Gornall, W. and Strebulaev, I. A., (2020) "Squaring Venture Capital Valuations with Reality". *Journal of Financial Economics*. Vol. 135, Issue 1, January 2020, Pages 120-143

Jarmulska, B. (2020). "Random Forest Versus Logit Models: Which Offers Better Early Warning of Fiscal Stress?". ECB Working Paper No. 2408 / May 2020.

Kaplan, S. N., B.A. Sensoy, and P. Strömberg. (2009). "Should Investors Bet on the Jockey or the Horse? Evidence from the Evolution of Firms from Early Business Plans to Public Companies." *Journal of Finance*, American Finance Association Vol. 64 (1): 75–115.

Keene, O. N. (1995). "The Log Transformation is Special". *Statistics in Medicine*. Vol. 14(8), Pages. 811-819.

Khan, I., Capozzoli, A., Corgnati, S. P., Cerquitelli, T. (2013). "Fault Detection Analysis of Building Energy Consumption Using Data Mining Techniques". *Energy Procedia*. Vol. 42, 2013, 557-566

Krzywinski, M., Altman, N. (2017) "Classification and regression trees". *Nature Methods*. Vol. 14, Pages 757–758. https://doi.org/10.1038/nmeth.4370

La Porta, R., Lopez de Silanes, F., Shleifer, A., and Vishny, R. (2002). "Investor Protection and Corporate Valuation", The Journal of Finance 57(3):1147-1170

La Porta, R., Lopez de Silanes, F., and Shleifer, A. (2002). "What Works in Securities Law", The Journal of Finance 61 (1): 1-32

Murray, G. R. and Scime, A. (2010) Microtargeting and Electorate Segmentation: Data Mining the American National Election Studies, *Journal of Political Marketing*, 9:3, 143-166, DOI: 10.1080/15377857.2010.497732

Nathan M. and Vandore E., (2014) "Here Be Startups: Exploring London's 'Tech City' Digital Cluster". *Environment and Planning A: Economy and Space*. Vol. 46(10): pages 2283-2299. doi:10.1068/a130255p

Porter, M. E. (1990) "The Competitive Advantage of Nations." *Harvard Business Review*. Vol. 68, no. 2 (March–April 1990): 73–93.

OECD (2019). "Financing SMEs and Entrepreneurs 2019. An OECD Scoreboard". Policy Highlights.

Popov, A. (2009) "Does Finance Bolster Superstar Companies? Banks, Venture Capital, and Firm Size in Local U.S. Markets", European Central Bank. Working Paper Series No. 1121.

Quintero, S. (2019). "Predicting a Startup Valuation with Data Science". *Journal of Empirical Entrepreneurship*. 2019.

Ribeiro-Oliveira, J. P., Garcia de Santana, D., Pereira, V. J., Machado dos Santos, C. (2018). "Data Transformation: An Underestimated Tool by Inappropriate Use". *Acta Scientiarum. Agronomy*. Vol. 40(1), e35300. https://doi.org/10.4025/actasciagron.v40i1.35300

Theall, K.P., Scribner, R., Broyles, S, Yu, Q., Chotalia, J., Simonsen, N., Schonlau, M., & Carlin, B.P. (2011). "Impact of small group size on neighbourhood influences in multilevel models". *Journal of Epidemiology and Community Health*, Vol. 65, pages 688–695.

Wooldridge, J. M. (2010). "Econometric Analysis of Cross Section and Panel Data". The MIT Press Cambridge, Massachusetts. 2010.

Zopounidis, C., Galariotis E., Doumpos, M., Sarri, S. and Andriosopoulos K. (2015) "Multiple Criteria Decision Aiding for Financial Decisions: An Updated Bibliographic Survey", *European Journal of Operational Research*, Vol. 247, pages 339-348.

# Appendix I: Example of CART-Based Microtargeting with One Categorical Variable

This appendix analyzes CART-based microtargeting approach to valuation using DCF-factors alongside the business-model categorical variable. In contrast to the OLS and CART regressions undertaken in the main sections of this paper, these are undertaken using the unmodified and untransformed dependent and independent variables, in order to yield a valuation-figure stated in Euros. Regression-tree models which do not use log-transformation have the advantage of expressing directly-actionable valuation-estimates, whose estimates are direct dollar-value estimations. On the other hand, this approach loses the outlier-mitigating model-flattening that occurs with log-transformation, making the model more susceptible to influence from extreme model-outliers. While this shortcoming can sometimes be mitigated using a random-forest approach, this has its limits when using categorical-variables, as there is a hard-limit on the number of categories random-forest can accommodate.

Compared to log-transformed regressions in the rest of the paper, here we see beta to be the primary branch of the regression tree, while firm-revenue is actually the most influential variable. In contrast with this table's log-transformed counterpart (i.e., Table 19), Table 23 assigns the primary deterministic impact-niche of customer-focus business-model to low-beta, but high-valuation startups, with B2B being indicative of a valuation-discount. Taken in conjunction with Table 19's findings, this indicates that the dataset's most substantial valuation-outliers are in low-beta sectoral-industries, and likely use B2C or B2B&C business-models.

*Table 23: Cart-based Microtargeting using one categorical-variable*

| OBS: | 1048 | | | |
|---|---|---|---|---|
| End Nodes: | 15 | | | |
| Complexity parameter | No. of Split | RMSE | Crossvalidation error | Crossvalidation St. Dev. |
| 0.1280 | 0 | 1.0000 | 1.0024 | 0.1634 |
| 0.0623 | 2 | 0.7441 | 0.8255 | 0.1484 |
| 0.0574 | 3 | 0.6817 | 0.8100 | 0.1479 |
| 0.0376 | 4 | 0.6243 | 0.7245 | 0.1398 |
| 0.0285 | 5 | 0.5867 | 0.7133 | 0.1397 |
| 0.0241 | 7 | 0.5296 | 0.7016 | 0.1385 |
| 0.0148 | 8 | 0.5055 | 0.6458 | 0.1366 |
| 0.0132 | 9 | 0.4906 | 0.6200 | 0.1317 |
| 0.0132 | 11 | 0.4643 | 0.6219 | 0.1318 |
| 0.0102 | 12 | 0.4512 | 0.6242 | 0.1318 |
| 0.0100 | 13 | 0.4409 | 0.6154 | 0.1318 |

Variable Importance

| Revenue | Business Model | Beta | Country-Risk Premium |
|---|---|---|---|
| 35 | 24 | 23 | 18 |

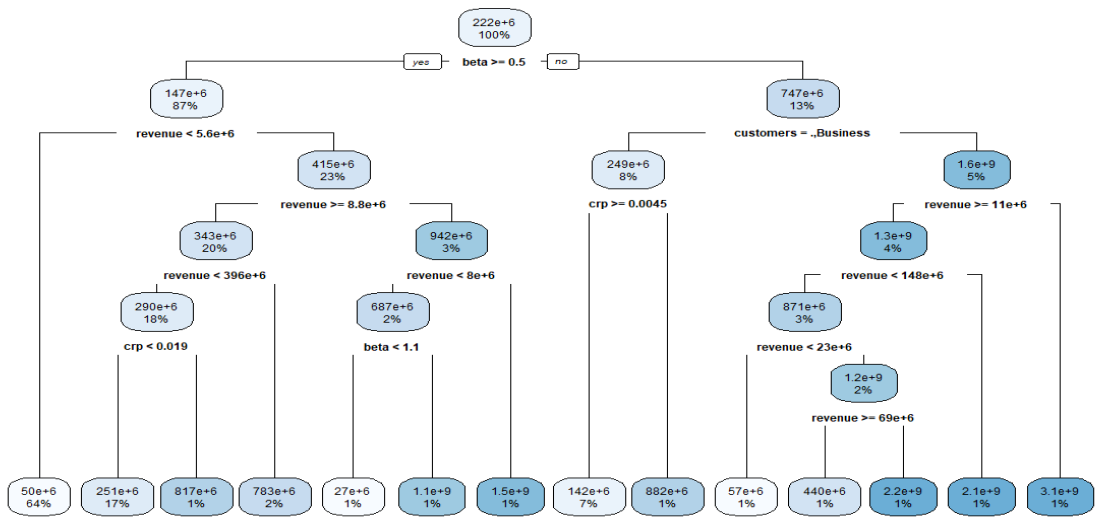

# Appendix II: Example of CART-Based Microtargeting with Multiple Categorical Variables

This appendix analyzes CART-based microtargeting approach to valuation using DCF-factors alongside country, industry-sector, business-model, and investor-type categorical variables. Regression trees which do not use log-transformation have the advantage of expressing directly-actionable valuation-estimates, whose estimates are direct dollar-value estimations. On the other hand, this approach loses the outlier-mitigating model-flattening that occurs with log-transformation, making the model more susceptible to influence from extreme model-outliers.

In contrast with this table's log-transformed counterpart (i.e., Table 21), Table 24 assigns city the highest importance, describing that municipal-geography plays a particularly-important role in higher-valuation startups. while assigning investor-type the lowest importance, with valuations being highly-sensitive to sectoral-beta in some cities, but to revenue in others.

*Table 24: Cart-based Microtargeting using multiple categorical-variables*

| OBS: | 1048 | | | |
|---|---|---|---|---|
| End Nodes: | 15 | | | |
| Complexity parameter | No. of Split | RMSE | Crossvalidation error | Crossvalidation St. Dev. |
| 0.1193 | 0 | 1.0000 | 1.0029 | 0.1637 |
| 0.1069 | 1 | 0.8807 | 0.9610 | 0.1549 |
| 0.0892 | 2 | 0.7738 | 0.9011 | 0.1508 |
| 0.0562 | 3 | 0.6845 | 0.7846 | 0.1429 |
| 0.0343 | 4 | 0.6283 | 0.6901 | 0.1341 |
| 0.0273 | 6 | 0.5597 | 0.6565 | 0.1317 |
| 0.0238 | 7 | 0.5324 | 0.6251 | 0.1251 |
| 0.0195 | 10 | 0.4611 | 0.5965 | 0.1174 |
| 0.0192 | 11 | 0.4416 | 0.5931 | 0.1175 |
| 0.0124 | 12 | 0.4224 | 0.6000 | 0.1175 |
| 0.0123 | 13 | 0.4100 | 0.5988 | 0.1195 |
| 0.01 | 14 | 0.3977 | 0.6013 | 0.1195 |

Variable Importance

| City | Beta | Sector | Country | Revenue | Business Model | Country-Risk Premium | Investor Type |
|---|---|---|---|---|---|---|---|
| 22 | 18 | 18 | 14 | 11 | 9 | 5 | 2 |

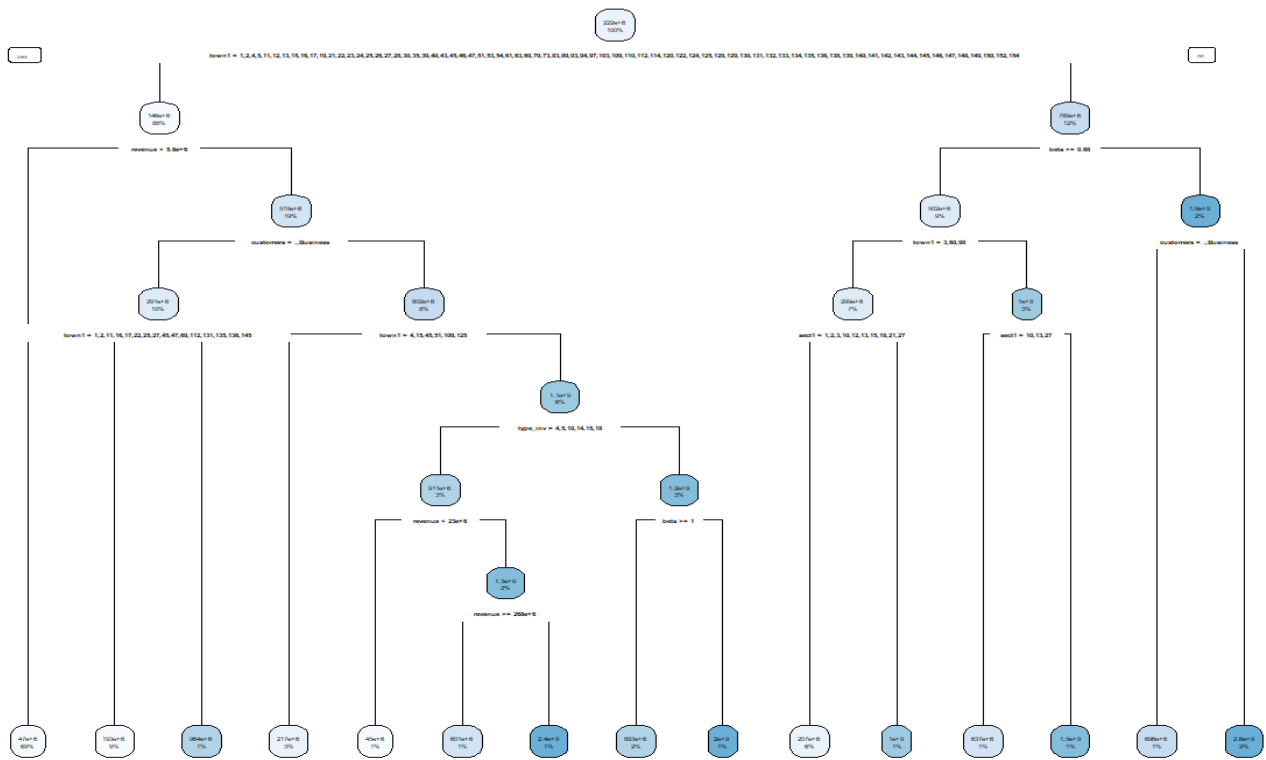